%% file: main.tex
\documentclass[nonacm, sigconf]{acmart}
\AtBeginDocument{%
  }

\setcopyright{none}
\renewcommand\footnotetextcopyrightpermission[1]{}
\acmISBN{978-1-4503-XXXX-X/2018/06}

\usepackage{graphicx}
\usepackage{wrapfig}
\usepackage{amsmath}
\usepackage{algorithm}
\usepackage{algpseudocode}
\usepackage{booktabs}
\usepackage{multirow}
\usepackage{enumitem}
\usepackage[table]{xcolor}
\usepackage{float}
\usepackage{dblfloatfix}
\usepackage{caption}
\usepackage{todonotes}
\usepackage{xspace}
\usepackage{cleveref}
\usepackage{amsthm}

\usepackage{listings}
\usepackage{xcolor}

\input{macros}

\lstdefinelanguage{TTL}{
  basicstyle=\ttfamily\footnotesize,
  morecomment=[l]{\#},
  morestring=[b]",
  sensitive=true,
  alsoletter={-,:}
}

\usepackage{tcolorbox}
\tcbuselibrary{breakable}
\usepackage{ragged2e}

\usepackage{subcaption} % for subfigures

\begin{document}

% \title{LoStBench: Benchmarking Longitudinal Studies Data Integration [Data-Intensive \& Data Science Application]}
\title{LoStBench: Benchmarking Longitudinal Studies Data Integration}

\author{Pratik Pokharel}
\orcid{1234-5678-9012}
\authornotemark[1]
\email{pratikpo@buffalo.edu}
\affiliation{%
  \institution{University at Buffalo}
  % \city{Buffalo}
  % \state{New York}
  \country{USA}
}

\author{Oliver Kennedy}
\email{okennedy@buffalo.edu}
\affiliation{%
  \institution{University at Buffalo}
  % \city{Buffalo}
  % \state{New York}
  \country{USA}
}
  
\author{Abhi Ramtel}
\email{aramtel@buffalo.edu}
\affiliation{%
  \institution{University at Buffalo}
  % \city{Buffalo}
  % \state{New York}
  \country{USA}
}
 
\author{Poonam Kumari}
\email{pkumari@bridgew.edu}
\affiliation{%
  \institution{Bridgewater State University}
  % \city{Bridgewater}
  % \state{Massachusetts}
  \country{USA}
}

 \author{Jeff Good}
\email{jcgood@buffalo.edu}
\affiliation{%
  \institution{University at Buffalo}
  % \city{Buffalo}
  % \state{New York}
  \country{USA}
}

  \author{Katarzyna Kordas}
\email{kkordas@buffalo.edu}
\affiliation{%
  \institution{University at Buffalo}
  % \city{Buffalo}
  % \state{New York}
  \country{USA}
}

 \author{Marianthi Markatou}
\email{markatou@buffalo.edu}
\affiliation{%
  \institution{University at Buffalo}
  % \city{Buffalo}
  % \state{New York}
  \country{USA}
}

  \author{Andrew Talal}
\email{ahtalal@buffalo.edu}
\affiliation{%
  \institution{University at Buffalo}
  % \city{Buffalo}
  % \state{New York}
  \country{USA}
}

\renewcommand{\shortauthors}{Pokharel et al.}
% \section{abstract}
\input{src/abstract}





\maketitle

\input{src/introduction}

\input{src/relatedworks}

\input{src/alignment}

\input{src/lostbench}

\input{src/goals}

\input{src/experiments}
\input{src/conclusion}

\bibliographystyle{ACM-Reference-Format}
\bibliography{references}

\end{document}

%% file: macros.tex
\newcommand{\xplural}{s\xspace}
\xspaceaddexceptions{\xplural}
\DeclareMathAlphabet{\mathbbold}{U}{bbold}{m}{n}

\newtoggle{shrinkspace}
\dimdef{\abovecapshrink}{-0.1cm}
\dimdef{\belowcapshrink}{-0.4cm}
\toggletrue{shrinkspace}

\newcommand{\trimfigurespacing}{\vspace*{\belowcapshrink}}

\newcommand{\intuple}[1]{\left\langle{#1}\right\rangle}
\newcommand{\comprehension}[2]{\left\{\;#1\;\middle|\;#2\;\right\}}

\newcommand{\abs}[1]{\left|#1\right|}
\newcommand{\schema}{S}
\newcommand{\targetschema}{\schema^T}
\newcommand{\attr}{a}
\newcommand{\mapfn}{\mathcal H}
\newcommand{\matching}{\mathcal M}
\newcommand{\gtmatching}{\matching^G}

\newcommand{\dist}{\textsc{dist}}
\newcommand{\similarity}{\textsc{sim}}
\newcommand{\arank}[2]{\textsc{arank}_{#1}(#2)}

\newcommand{\mrank}[2]{\textsc{mrank}(#1, #2)}
\newcommand{\attrprop}[1]{\textsc{#1}}
\newcommand{\categoryset}{\mathcal{C}}
\newcommand{\cat}{c}
\newcommand{\ragt}{\text{R@GT}}
\newcommand{\mrr}{\text{MRR}}
\newcommand{\rir}{\text{RIR}}

%% file: src/abstract.tex
\begin{abstract}
A longitudinal study observes the same subjects or variables, repeatedly over an extended period of time, allowing researchers to examine changes, trends, and relationships as they develop.
Such studies typically evolve over their life cycles, as improved understanding of the study's domain, or external factors trigger the need for revisions.
The resulting heterogeneity among collected data must be resolved before the survey's results may be analyzed.
Although this problem is a form of schema matching, it is a form of schema matching that differs significantly from prior workloads explored by the database community.
For instance, rather than the short attribute names and data instances that many schema matching tools rely on, a stronger signal in longitudinal studies arises from longer prose questions that these tools are less equipped to handle.

This paper documents the joint efforts of a team of computer scientists and domain experts to adapt existing schema matching algorithms to our new use case of longitudinal data integration.
We present a case study in the context of a representative workload: The American National Election Studies (ANES) dataset, a nearly eight-decade survey with publicly available data and a canonical domain-expert curated schema matching.
Using ANES, we develop \textbf{LoStBench} (\textbf{Lo}ngitudinal \textbf{St}udies \textbf{Bench}mark for data integration), a medium-scale, subject-matter-expert–annotated benchmark.
Using LoStBench, we conduct a comprehensive empirical study evaluating state-of-the-art schema matching algorithms, and expose limitations of these algorithms when applied in the context of longitudinal studies.
We propose improvements that address some of these limitations, and identify key challenges to motivate future data integration work to address the others.
\end{abstract}

%% file: src/introduction.tex
\section{Introduction}

A longitudinal study observes the same subjects or variables, repeatedly over an extended period of time, allowing researchers to examine changes, trends, and relationships as they develop.
Survey-based longitudinal studies, where researchers poll participants using a questionnaire, or "survey instrument" arise in a wide range of research disciplines.
A common feature of longitudinal studies is that they evolve over time.
For example, revisions to the study may be triggered when researchers develop a deeper understanding of the study domain that renders a question invalid or ambiguous; or when decades-long shifts in cultural norms affect the interpretation of a question.
% (iii) external factors (e.g., conflict) affect the study population and require adjusting data collection practices;
% (iv) external factors (e.g., inflation) affect the context of collected data and require re-interpreting answers.
As a survey instrument is revised, the data collected across multiple revisions becomes heterogeneous; each revision of the study produces a dataset with a distinct schema.
To analyze the resulting data, researchers expend significant manual effort integrating data collected across the survey revisions.

\begin{figure*}[t]
    \centering 
    \includegraphics[width=0.81\linewidth]{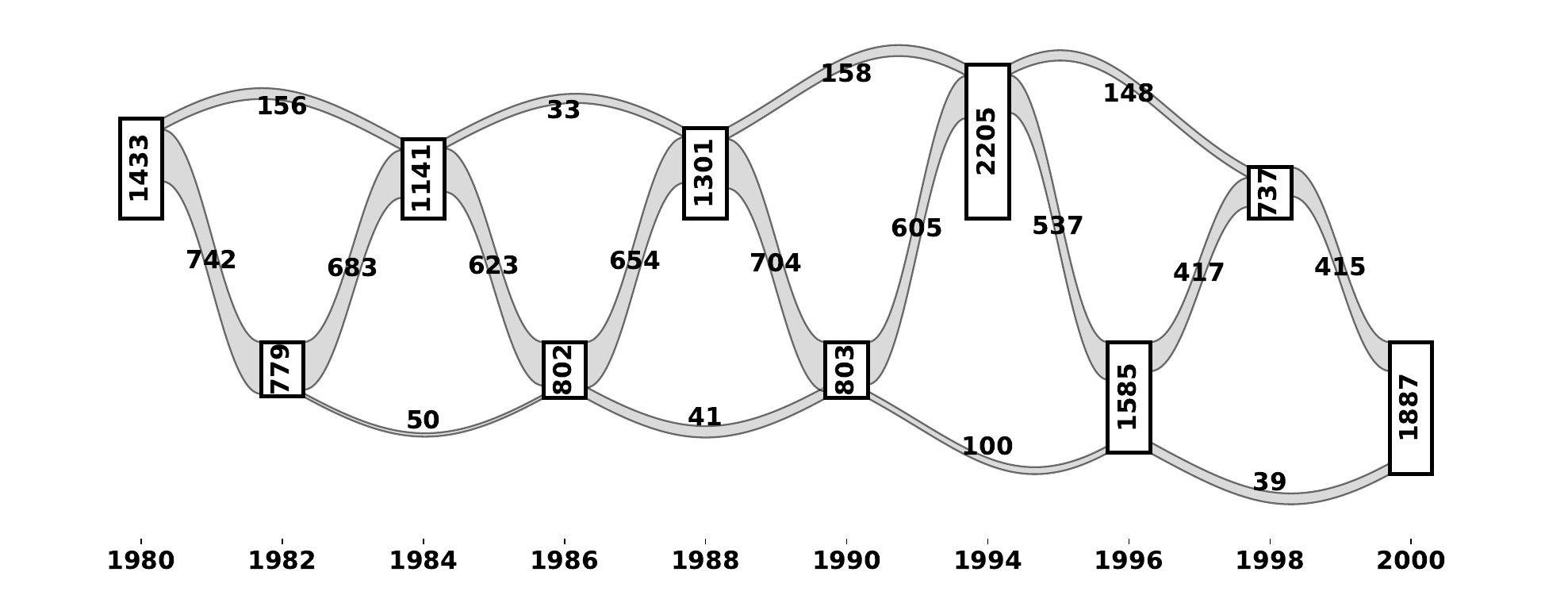}
    \caption{A sample of flow of related variables across years in the ANES dataset \cite{anes2026_codebook}. Rectangular Nodes are dataset versions annotated with the number of variables in that version. We show the flow between adjacent and every second versions only for clarity in the figure. The curved flow joining two adjacent versions represent the number of variables present in both the versions. ANES alternates pre- and post- election questionnaires, and so the remaining lines link surveys of the same type.
    }
    \label{fig:anes_dataflow}
    \trimfigurespacing
\end{figure*}

This is, at first glance, a classical data integration problem: We want to map a collection of datasets with distinct (local) schemas into a uniform (global) schema.
More concretely, this is a problem of \emph{schema matching}, or establishing correspondences between attributes in related schemas.
Schema matching is addressed as part of data integration~\cite{data-integration_textbook, data_exchange_Marcelo_Arenas_textbook, hetero_data_integration_sonia_bergamaschi, data_int_and_engg_general_article}, table enrichment~\cite{3.fatemeh} and semantic type detection~\cite{sherlock}.
There is potential in existing tools to be a good fit for longitudinal studies as well.

The authors of this paper include experts from diverse set of disciplines, with experience conducting longitudinal studies in the fields of linguistics, environmental health, and medicine.
This paper documents our ongoing\footnote{Or as one might say, longitudinal} effort to adapt existing schema matching techniques into the context of longitudinal studies.
To focus this discussion, and to make it easier for others to build on our efforts, we implemented a case study using a dataset from The American National Election Studies (ANES)\cite{anes2026_codebook}, a longitudinal study that has been running for nearly eight decades.
ANES is ideal, because (i) it captures many of the challenges that we have faced in our own longitudinal data integration efforts, and (ii) its codebook defines a hand-curated ground truth schema matching.
% Concretely, this paper presents the Longitudinal Study Benchmark (LoStBench), a benchmark based on the ANES schemas, datasets, and codebook\footnote{Source code and Data repo: \url{https://anonymous.4open.science/r/lostbench-AFCC/}}.
Concretely, this paper presents the Longitudinal Study Benchmark (LoStBench), a benchmark based on the ANES schemas, datasets, and codebook\footnote{Source code and Data repo: \url{https://github.com/UTDRON/lostbench}}.
We explore and address the challenges of adapting algorithms for schema matching and related problems, both to ANES specifically, and to longitudinal studies in general, and then evaluate a wide range of existing algorithms on this new and rather distinct use case.

Although the ANES data and schema share many statistical properties with common schema matching benchmarks, several key differences affect the effectiveness of existing matching algorithms.
First, while ANES provides attribute names for each of its schemas, the prose of the questions provides a far richer signal to mine for attribute matchings.
We found that schema matching algorithms that can effectively incorporate this prose, for example via word embeddings, to be substantially more effective.
Similarly, compared to other schema matching problems, ANES features a relatively low level of data variation due to common, repeating question patterns, as well as a rather high null response rate due to unanswered questions, or frequent use of schema denormalization.
As a result, we found that instance-based techniques perform more poorly than metadata-based techniques.
Finally, some iterations of the ANES ask as many as 2000 questions.
Such wide schemas pose scalability problems for schema integration algorithms intended for use with dozens, or at most hundreds of attributes.
We propose a blocking-style approach to schema integration based on table unionability.
Finally, we propose and evaluate an algorithm to leverage the sequential nature of longitudinal datasets.

%%%%%%%%%%%%%%%%%%%%%%%%%%%%%%%%%%%%%%%%%%%%%%%%%%%%%%%%%%%%%%%%%%%%%%%%%%%%%%%%%%%%%%
%%%%%%%%%%%%%%%%%%%%%%%%%%%%%%%%%%%%%%%%%%%%%%%%%%%%%%%%%%%%%%%%%%%%%%%%%%%%%%%%%%%%%%
\input{src/section2}

%%%%%%%%%%%%%%%%%%%%%%%%%%%%%%%%%%%%%%%%%%%%%%%%%%%%%%%%%%%%%%%%%%%%%%%%%%%%%%%%%%%%%%
%%%%%%%%%%%%%%%%%%%%%%%%%%%%%%%%%%%%%%%%%%%%%%%%%%%%%%%%%%%%%%%%%%%%%%%%%%%%%%%%%%%%%%
\subsection{Overview}
The balance of this paper is organized as follows.
In \Cref{sec:preliminaries}, we formally define the problem of schema matching and put past work in schema matching and similar problems into a shared context.
\Cref{sec:attribute-alignment} presents our approach to adapting schema matching algorithms to longitudinal data integration.
\Cref{sec:case-study} and \Cref{sec:long-data-int-goal} outline LoStBench, first by reviewing the ANES dataset, and then our experimental goals for the benchmark.
\Cref{sec:experiments} presents the results of our evaluation, and we conclude in \Cref{sec:conclusion}.

%% file: src/section2.tex
\subsection{Longitudinal Data Management: In Practice}
\label{sec:inpractice}

The data collection and management problems of longitudinal studies are relevant to most research that involves collecting data on human social behavior, especially outside of highly controlled settings.
They are especially common for surveys and interviews, when a researcher enters data based on interviewee responses, or when the data is entered directly by study participants.
Among the authors of this paper are researchers from three distinct domains: linguistics, environmental health, and healthcare.
% \footnote{Names and identifying project details are elided to preserve double-blindness}.
Although we use ANES as a representative use case for our benchmark due to its publicly available ground truth, we motivate the need for data integration techniques specialized for longitudinal data by outlining concrete challenges that we have experienced.

\paragraph{Linguistic Research}
Linguistic research on underdescribed languages typically involves a linguist interacting with users of an underdescribed language to gather data about the language and its users~\cite{ChelliahDeReuse:2011,ThiebergerBerez:2012,Good:ScopeData}.
Alongside recordings of both structured interactions and naturalistic examples of language use, sociolinguistic interviews are often used to gather information about the linguistic background of individuals from whom data has been collected. A questionnaire or similar instrument is often used to conduct these interviews, and they are administered by members of the research team.
One of our authors has been conducting a multi-year linguistic study in Cameroon, which has required multiple revisions to the data collection protocol. 
Some changes are necessary to cope with exogenous factors like political conflict: e.g., ``What village are you from?'' becomes ``What village did you live in before being displaced?''
Conversely, other changes result from improved domain understanding: e.g., A discovery that some individuals use different names when interacting with different people required the survey to be adjusted to allow for an individual to provide multiple names and the contexts in which they use them.

\paragraph{Environmental Health Studies}

Pollution in an individual's environment is frequently correlated with social factors like the individual's role in the family (e.g., male vs. female caregiver), income, resources or social position, or the attributes of the neighborhood where the family resides.
% Intersectionality may exist among race/ethnicity, gender, and social capital, as well as regional or country-wide policies.
% For example, there is strong evidence of environmental injustices occurring at many levels of society, with underrepresented or socially disadvantaged racial/ethnic groups being more exposed to environmental pollution than their economically privileged, white, neighbors.
% Similarly, household-level use of clean or solid fuels for cooking/heating is related to the proportion of rural populations in the country as well as the country's development index and economic incentives offered to users.
Another of our authors has been directing a research program in Montevideo, Uruguay since 2009, collecting data on individual, family, and neighborhood-level factors relating to exposure metrics and health effects.
The current round of data collection, with annual surveys between 2021 and 2024, involves 15 survey instruments posed to caregivers and youth totaling over 250 questions.
An evolving understanding or operationalization of social factor variables (e.g., differential income distribution due to inflation or the changing availability of different types of items or possessions like smartphones that may be used to calculate asset-based household socioeconomic status score), and changes in exposure patterns and/or definitions (e.g., better understanding of the exposure sources or patterns for different environmental chemicals), have necessitated changes in the wording of questions, including question removal, addition, and alteration. 
Over time, survey respondents evolve as well:
Initially, when study participants were children, the caregivers served as primary study informants.
Over time, responding has shifted to the youth, who can report on their own behaviors or states (e.g., mental health). 
These shifts require appropriate tools to annotate, combine, and analyze individual question responses as well as any derived indices.

\paragraph{Treatment for Hepatitis-C}
Social, behavioral, and environmental factors contribute to health inequalities and play a disproportionately large role in health outcomes.
Such factors contribute substantially to an individual's overall physical and mental health, and can have a deleterious effect on health outcomes~\cite{Daniel2018}.
Specifically, low literacy, racial segregation, poverty, low socioeconomic support, and low levels of social support contribute substantially to mortality~\cite{Galea2011}.
Recently, professional societies, such as the American College of Physicians (ACP), have highlighted research gaps in the area of SDOH based upon the realization that they require prioritization in order to improve health outcomes, particularly among disadvantaged and under-served populations~\cite{Daniel2018,Taylor2016,ThomasHenkel2017}.
A third author is working on a project to understand the relationship between substance abuse and social factors, as part of research on improving treatment for Hepatitis-C.
A key source of data for this project is patient in-take forms from methadone clinics collected across multiple care providers, and over the course of decades.
Each care provider asks a different, though often related, set of questions.
Moreover, even within a single care provider, the structure of the forms evolves over time, with questions added, removed, and/or rephrased.

%% file: src/relatedworks.tex
\section{Preliminaries}
\label{sec:preliminaries}
% S3. Defining the Alignment problem and a brief discussion of related work in the DB community

Data integration concerns combining data from multiple heterogeneous sources to provide a unified view that can serve as a foundation for downstream analytics~\cite{data_integration_theoretical_perspective}.
Despite its seemingly straightforward definition, data integration involves significant challenges, including system-level issues, logical heterogeneity across sources, and administrative barriers such as siloed data ownership~\cite{data-integration_textbook, di_challenges}.

While integration objectives may vary depending on the workload, a fundamental subtask is \emph{schema matching}, the identification of correspondences and relationships across datasets and their attributes.
Schema matching is a core component of data integration and arises in numerous applications, including data warehousing, e-business, and semantic query processing~\cite{a_survey_of_sm_approaches}.
In such a task, we are generally given 2 ``source'' dataset instances with schemas $\schema_1$, $\schema_2$, where each schema $\schema_i$ is a set of attributes $\attr \in \schema_i$.
We are told that these datasets correspond to the same logical entity, and so we would like to connect the attributes of the two schemas.

Naively, we can construct a unified ``target'' dataset with schema $\targetschema$, and then establish partial mapping functions $\mapfn_i : \schema_i \rightarrow \targetschema$ from the source schemas to the target.
However, the target schema is often unknown at first, and so an equivalent framing of the problem is to discover a set of pairwise attribute matchings $\matching$ such that every bi-clique in $\matching$ corresponds to a target attribute.
$$\matching = \comprehension{\intuple{\attr, \attr'}}{\attr \in \schema_1, \attr' \in \schema_2, \mapfn_1(\attr) = \mapfn_2(\attr')}$$
Schema matching algorithms infer $\matching$ by leveraging properties of each attribute, which may include schema-level \emph{metadata}~\cite{coma, sim_flooding, 1.DDM, cupid, infogather}, instance-level data~\cite{distr_based, infogather}, or a combination of the two~\cite{magneto, coma++, valentine}.
Regardless of the specific properties leveraged, the typical approach is to define an easily computable distance measure\footnote{or equivalently a similarity measure} over pairs of attributes $\dist(\attr, \attr')$.
Ideally, for the ground truth matching $\gtmatching$, lower values of the measure should correlate with a greater chance that $\intuple{\attr, \attr'} \in \gtmatching$ for a given pair $\attr \in \schema_i$, $\attr' \in \schema_j$.
When aligning schemas, we are additionally interested in two ranking schemes derived from the distance measure.

\begin{definition}[Attribute Rank]
Given an attribute $\attr \in \schema_1$, the attribute rank, denoted $\arank{\attr}{\attr'}$ is defined with respect to the list of attributes $\attr' \in \schema_2$ that could potentially match with $\attr$:
$$\arank{\attr}{\attr'} = \abs{\comprehension{\attr''}{\dist(\attr, \attr'') \leq \dist(\attr, \attr')}}$$
\end{definition}

\begin{definition}[Matching Rank]
  \label{def:mrank}
  The rank of a potential matching $\intuple{\attr, \attr'}$ is given by the list of all matches ordered by distance:
$$\mrank{\attr}{\attr'} = \abs{\comprehension{\intuple{\attr'', \attr'''}}{\dist(\attr'', \attr''') \leq \dist(\attr, \attr')}}$$
\end{definition}

We refer to the set of all matchings of rank $k$ or below as the ``top-$k$'' candidate matching.

\paragraph{Direct Schema Matching}
Early methods relied on surface-level similarities like lexical overlap or distributional similarity using heuristics that limited their ability to capture deeper semantic relationships.
An earlier work \cite{ml_for_sm} framed schema matching as a machine learning classification problem using metadata and data features, showing that learned models outperform traditional heuristic methods.
Now, some more recent works capture deeper semantics using embeddings~\cite{semprop} and pretrained language models~\cite{1.DDM, magneto, unicorn}.
In this paper, we evaluate the distance functions of the following schema matching techniques.

\begin{enumerate}[leftmargin=0.0in,label=]
    \item \textbf{Cupid \cite{cupid}} represents schemata as trees to capture the hierarchy of elements (relations, attributes, etc.). Similarity between elements is computed as a weighted combination of linguistic (name) and structural (contextual) similarities.
    \item \textbf{Similarity Flooding \cite{sim_flooding}} transforms schemas into directed graphs and merges them into a propagation graph. Similarities propagate iteratively between connected elements until convergence.
    \item \textbf{COMA/COMA++ \cite{coma,coma++}} models schemas as directed acyclic graphs and combines multiple matchers (schema-based and instance-based).
    \item \textbf{Distribution-Based Matching \cite{distr_based}} computes similarities between columns based on the distribution of their data values. It clusters relational attributes using the Earth Mover’s Distance and outputs disjoint clusters of related attributes.
    \item \textbf{Jaccard-Levenshtein Matcher}- a simple baseline that computes pairwise column similarity using Jaccard similarity, treating values as identical if their Levenshtein distance falls below a threshold.   
\end{enumerate}

\paragraph{Ontology Based Matchers}
Ontology-based schema matching algorithms like \textsc{SemProp}~\cite{semprop} rely on a having domain-specific ontology.
To our knowledge, a low level ontology for election concepts or survey variables in general is not available.
Many attributes in LoStBench correspond to literal responses, rather than the ontology classes or instances used in general ontologies like DBpedia \cite{dbpedia} and YAGO \cite{yago_kb}.
Conversely, existing survey specific ontologies like the survey ontology~\cite{survey_ontology} are intended as frameworks into which an ontology might be built.
As a result, ontology-based methods such as SemProp are not directly applicable.

\paragraph{Schema Matching with Re-ranking}
With the emergence of large language models (LLMs), several studies have explored prompt-based schema matching, highlighting the sensitivity of performance to the amount and structure of context provided~\cite{sm_with_llm_monolithic}.
LLM-based schema matching techniques adopt multi-stage pipelines \cite{magneto, construm, 1.DDM} to optimize LLM usage; for example, 
% Magneto~\cite{magneto} first produces a ranked list using a pretrained language model and then refines it via LLM-based reranking, while 
Construm~\cite{construm} enhances decision quality by injecting structure-aware context into candidate selection or re-ranking under constrained budgets.
In this work, we evaluate the re-ranking ability of two such systems:

\begin{enumerate}[leftmargin=0.0in,label=]
    \item \textbf{Magneto \cite{magneto}} leverages both small and large language models. It first retrieves candidate matches using a self-supervised, fine-tuned small LM (SLM), then re-ranks them with an LLM.
    \item \textbf{DDM \cite{1.DDM}} uses sentence embeddings of prose questions from pretrained language models and then performs approximate nearest neighbor search to retrieve ranked candidate lists.  It is our own previous work and the version we evaluated optionally re-ranks attributes using an LLM.
    % On an extension, we also ran an LLM re-ranker over its output. 
\end{enumerate}

\paragraph{Table Enrichment}
Attribute alignment also arises in the context of table enrichment problems like dataset discovery~\cite{finding_related_tables_alon_halevy, aurum}, where the goal is to identify tables that are unionable (adding tuples) or joinable (adding attributes) with respect to a query table.
That is, given some query schema $\schema^Q$ and a repository of schemas $\schema_1, \ldots, \schema_N$, the goal is to select the $k$ most "similar" schemas.
In contrast to schema matching, the distance (similarity) measure is defined at the level of schemas, rather than individual attributes.
However, a common approach is to define the distance between the two schemas in terms of a potential schema matching.
For example, Nargesian et~al.~\cite{3.fatemeh} propose a measure (c-alignments) in which the distance between two schemas is computed by proposing a schema matching between them, and then scoring the matching based on a distance measure.
That is, given a computed matching $\matching$ for schemas $S_i$, $S_j$, we can define the similarity as:

\begin{equation}
  \label{eqn:similarity}
  \similarity(i,j) = \sum_{\intuple{\attr, \attr'} \in \matching_{i,j}} (1-\dist(\attr, \attr'))
\end{equation}

% \begin{definition}[Unionable Schema Rank]
% Given a schema $\schema $, the unionable schema rank, denoted by $\srank{\schema}{\schema'}$ is defined with respect to the list of schemas $\schema' \in \bigcup_i\schema_i\setminus\schema$ that could potentially union with $\schema$:
% $$\srank{\schema}{\attr'} = \abs{\comprehension{\attr''}{\dist(\attr, \attr'') \leq \dist(\attr, \attr')}}$$
% \end{definition}

Crucially, even though the objective of table enrichment systems is inter-schema similarity, they are nevertheless frequently implemented in terms of inter-attribute distance measures, and can be adapted for use in schema matching.
In this paper, we evaluate the following table unionability systems:

\begin{enumerate}[leftmargin=0.0in,label=]
    \item  \textbf{SANTOS \cite{4.aamod}} performs table union search using semantic relationships between column pairs, beyond traditional schema or value-based similarity. It leverages external knowledge bases and synthesized knowledge derived from data lakes to find matchings.
    
    \item \textbf{Starmie \cite{5.fan}} formulates table union search as a representation learning problem by training column encoders using contrastive learning over table data. It computes unionability via cosine similarity between learned column embeddings and employs efficient indexing structures such as HNSW for fast retrieval.

    \item \textbf{D3L \cite{d3l_bogatu}} approaches table union search by extracting name-, value-, format-, embedding-, and distribution-based dataset features and mapping them into a uniform distance space using hash-based indexing. Similarity between tables is computed based on distances between these feature representations, enabling efficient retrieval of related tables.
    
    \item \textbf{MetaTUS \cite{2.martorana2025metadatadriventableunionsearch}} performs table union search using only metadata, avoiding reliance on raw data values to address privacy constraints. It models unionability via schema-level and descriptive metadata features, enabling integration under restricted data access while still aligning with FAIR principles \cite{margherita_fair}. 
    
    % \item \textbf{DDM \cite{1.DDM}} is also a metadata only method like MetaTUS. It uses sentence embeddings for prose questions and performs efficient similarity search to find similar variables, then performs a greedy bipartite matching among the schemas, whose aggregate scores are then used to rank the unionable tables.
    % The primary purpose of DDM is schema alignment, but we extended its functionality by performing a greedy bipartite matching among the schemas, whose aggregate scores were then used to rank the unionable tables.
\end{enumerate}

Other works we were not able to obtain for evaluation include TACTUS~\cite{tactus}, which introduces a table-centric approach that first retrieves candidates using table-level embeddings and subsequently refines matches at the column level, improving both efficiency and ranking quality.
Complementary to union search, join discovery methods~\cite{nested_data_search, josie, high_dim_join_discovery} focus on extending tables horizontally.
% Despite substantial progress in the context of table enrichment, instance-driven attribute alignment faces inherent limitations when data values are numerical or lack explicit semantic cues.
% Such approaches are also of limited use when the instance is not available, for example due to privacy concerns.
% Metadata-driven approaches offer a practical alternative in such settings.
% MetaTUS~\cite{2.martorana2025metadatadriventableunionsearch}, for instance, leverages metadata aligned with FAIR principles~\cite{margherita_fair} and incorporates LLM based re-ranking to further enhance performance.
Evaluation of table union search methods is supported by diverse benchmarks, including LLM generated datasets~\cite{alt_gen}, human-annotated corpora~\cite{pylon_tus}, and synthetic benchmarks ~\cite{lakebench, lakebench_aamod, 3.fatemeh, 4.aamod} constructed by decomposing large seed tables.

\subsection{Other Related Work}
\paragraph{Table Representation}
Closely related to the above two tasks is table representation learning, which underpins a wide range of tabular data applications.
Prior work explores different objectives, including table understanding (TURL~\cite{TURL}), question answering (TABBIE~\cite{tabbie}), and semantic type detection~\cite{dosolo_doduo, sherlock, sato}.
Notably, several methods for schema matching and table union search~\cite{2.martorana2025metadatadriventableunionsearch, 5.fan, valentine, unicorn} build upon similar representation techniques.
However, learning meaningful representations from data instances remains challenging, particularly on high-entropy numerical data~\cite{sportstables}.
To address this, Pythagoras~\cite{pythagoras} proposes a graph based formulation of tables combined with graph neural networks, demonstrating improved performance in capturing semantics.

\paragraph{Merging Data Models}
Most schema matching approaches focus on pairwise alignment; however, real-world data integration scenarios often involve multiple schemas, leading to the \emph{schema matching network} problem~\cite{corpus_based_sm_network, hollistic_sm, pay_as_you_go_sm}.
Rodrigues et. al \cite{ml_techniques_for_sm_network} addresses this problem of schema matching in multi-schema environments by proposing machine learning based technique that leverage network constraints and bootstrapped training data.
In such settings, curators must identify correspondences across many schemas simultaneously, a task with combinatorial complexity.
This challenge is especially pronounced in longitudinal studies, where multiple schema versions must be aligned and consolidated into a unified representation.
So, beyond identifying correspondences, an equally important task is schema merging: given multiple schemas and their mappings, the goal is to construct a duplicate free and conflict resolved global model.
Early work by Buneman ~\cite{buneman1992schema_merging_theoretical} provides a theoretical foundation by modeling schemas as graphs and defining merge operations based on information ordering.
Pottinger and Bernstein~\cite{pottinger2003merging} propose a general framework for resolving duplicates and conflicts during merging.
Other approaches focus on practical interoperability across heterogeneous models; for example, Quix et al.~\cite{quix2007generic_merge} introduce a metamodel based solution that transforms schemas into a metamodel, followed by human assisted mapping using real world semantics prior to merging.

When extending beyond two schemas, the merging process becomes more complex due to ordering effects and accumulated inconsistencies.
To address this, Boubakir and Chaoui~\cite{boubakir2016pairwise_merge_multiple_models} propose a similarity-driven, iterative pairwise merging strategy that prioritizes merging closely related schemas to improve overall quality.

\paragraph{Integrating Longitudinal Datasets}
Prior work on integrating longitudinal datasets remains relatively limited.
Many studies rely on manual curation supplemented with rule-based semi-automated techniques for interoperability across datasets \cite{mental_health_OMOP, poverty_panel_study, multiomics_obesity}.
For example, in a panel study on poverty \cite{poverty_panel_study}, person-level record linkage is performed using deterministic, rule-based matching strategies.
In healthcare domains, integration efforts frequently focus on aligning schemas to predefined standards or ontologies \cite{mental_health_OMOP, 10.3389/fmed.2023.1076794}, or transforming data instances into structured mathematical representations \cite{multiomics_obesity}.
Additionally, some works address construct level alignment by fusing heterogeneous measurement scales across different datasets to enable unified modeling \cite{measurement_scale_fusion}.

% \pk{following section may be unhelpful}
% A separate line of research focuses on multimodal data integration at the feature or representation level, rather than explicit schema or instance alignment.
% These approaches aim to construct unified representations from heterogeneous data sources.
% For instance, \cite{feature_fusion_medical_long_studies} combines features from clinical, imaging, and genomic data to engineer an integrated feature for an enhanced predictive modeling.
% Similarly, \cite{multiomics_integration} integrates longitudinal multi-omics datasets into a unified tensor representation.
% \pk{till here}
In a related direction, \cite{long_cross_merge_get_pseudo_time_series} integrates cross-sectional and longitudinal data by generating pseudo time-series from the former and calibrating them using a small number of longitudinal samples to support disease progression modeling.
Other works, such as \cite{long_study_growth_faltering_preterm_infants}, integrate structured and unstructured electronic health record data by linking patient records across modalities using deep learning techniques.
% \pk{I feel the whole paragraph can be omitted based on reading just section 2}

Recent approaches have also explored the use of pre-trained language models to support longitudinal schema alignment.
These methods leverage the semantic content of survey instruments or variable descriptions, rather than relying on data values, to identify candidate correspondences \cite{10.3389/fmed.2023.1076794,1.DDM}.
Typically, similarity-based matching or clustering is used to propose alignments, which are subsequently validated and refined by domain experts to ensure semantic consistency. 

Finally, some approaches do not directly integrate datasets, but instead combine insights derived from longitudinal data.
These include ensemble frameworks that combine predictions from multiple longitudinal modalities using models such as LSTMs \cite{long_multimodels_separate_predictions_combinations}, as well as methods that incorporate features or latent representations from longitudinal data into analysis models for improved predictions \cite{nguyen_paris_survival_analysis}.
Related efforts also explore knowledge-level integration by linking survey questionnaire concepts through knowledge graphs, enabling adaptive systems that dynamically tailor questions based on user responses \cite{knowledge_graph_social_survey}.

%% file: src/alignment.tex
\section{Attribute Alignment for Longitudinal Studies}
\label{sec:attribute-alignment}

In contrast to basic schema matching settings, where the focus is on integrating pairs of schemas, longitudinal datasets consist of \emph{sequences} of schemas $\schema_1, \ldots, \schema_N$.
Ultimately the goal is the same: constructing a single target schema $\targetschema$ by discovering inter-attribute mappings across the entire schema.
We denote the set of mappings between schemas $\schema_i$ and $\schema_j$ as $\matching_{i,j}$.

Pairs of successive schemas $\schema_i$, $\schema_{i+1}$ frequently have significant overlap.
That is, for many distance measures, the distance between many attribute pairs in successive schemas is negligible (i.e., given $\intuple{\attr, \attr'} \in \matching_{i, i+1}$, w.h.p. $\dist(\attr, \attr') \approx 0$).
We leverage this fact through an \emph{iterative} integration strategy, where the target schema is incrementally developed by rolling up one schema at a time, in sequence.
This strategy is summarized in \Cref{alg:incr-integ}.

\begin{algorithm}
\caption{IncrementalIntegration($\schema_1, \ldots, \schema_N$)}
\label{alg:incr-integ}
\begin{algorithmic}
    \Require{$\schema_1, \ldots, \schema_N$: A sequence of schemas}
    \Ensure{$\targetschema$: The target schema, \hfill $\mapfn_1, \ldots, \mapfn_N$: Mappings to $\targetschema$}
    \State{$\targetschema \gets \schema_1$; \textbf{define} $\mapfn_1(\attr) = \attr$}
    \ForAll{$i \gets 2 \ldots N$}
        \State{\textbf{compute} $\matching_i$ \textbf{between} $\targetschema$ \textbf{and} $\schema_i$}
        \State{$\targetschema \gets \targetschema \cup \left(\schema_i - \comprehension{\attr}{\intuple{\attr,\_}\in \matching}\right)$}
            \Comment{Add unmatched attrs.}
        \State{\textbf{define} $\mapfn_i(\attr) = \begin{cases}
            \attr' & \textbf{if } \intuple{\attr', \attr} \in \matching_i\\
            \attr & \textbf{otherwise}
        \end{cases}$}
    \EndFor
\end{algorithmic}
\end{algorithm}

We note several subtle details about this algorithm.
First, when computing each matching $\matching_i$, we have access to metadata for every preceding attribute linked into $\targetschema$; Thus exact matches can be constructed immediately.
Second, in practice, matching heuristics are imprecise, and so human direction is frequently critical.
We note that the process of constructing matchings is an ideal point for human intervention, as it closely mirrors the iterative nature of incremental data releases for long running projects like ANES.

\subsection{Wide Table Support}
\label{sec:wide-table}
Survey instances in our representative use case of ANES are quite wide (discussed in detail in \Cref{lostbench_tables_synthesis}), with 100 to 2200 attributes per version.
Such large schemas are common to many intensive surveys of this type, and so to be useful for longitudinal studies, schema matching tools must support wide schemas.
Unfortunately, not all of them presently scale to schemas with thousands of attributes.
For example, MagnetoGPT \cite{magneto} is limitated by its LLM-based re-ranking step, where long candidate lists (that too containing long prose questions) quickly exceed prompt limits, or degrade performance due to context loss.
Even metadata-based methods like Cupid \cite{cupid} do not scale well because it performs exhaustive pairwise comparisons and repeatedly refines matches using structural relationships in the schema.
Similarly, SANTOS \cite{4.aamod} functional dependency computation step can take in excess of 24 hours for as few as 200 attributes.

Fortunately, large surveys like ANES follow a hierarchical structure that gives us a way to scale attribute-limited alignment schemes.
ANES attributes are clustered into groupings of thematically related questions (denoted $\attrprop{category}(\attr)$) that correlate with schema matchings.
When aligning two schemas, we block~\cite{baxter2003blocking} categories through a technique similar to table union search \cite{3.fatemeh}.
Let the category set of schema $\schema_i$ be
$\categoryset_i = \bigcup_{\attr \in \schema_i} \attrprop{category}(\attr)$
For each category $\cat \in \categoryset_i$, we vertically partition $\schema_i$ into sub-schemas:
$$\schema_{i,\cat} = \comprehension{\attr}{\attr \in \schema_i \wedge \attrprop{category}(\attr) = \cat}$$

Recall that we can compute the similarity between pairs of schemas $\schema_{i, \cat}$, $\schema_{j,\cat'}$ with \Cref{eqn:similarity} by performing a schema match over the two schemas.
For each $\schema_{i, \cat}$, we find the $\cat' \in \categoryset_j$ that maximizes $\similarity(\schema_{i, \cat}, \schema_{j, \cat'})$ and take the union of all such matchings.
This approach is quadratic in the number of categories, but like blocking in record linkage~\cite{baxter2003blocking}, avoids being quadratic in the attributes.

% To construct test cases suitable for all the models used in our experiments, we generated tables by grouping variables according to their categories $\mathcal{C}_i$. For each category $c_{ik}$ within survey year $y_i$, we created a table $\mathcal{T}_{ik}$ because variables belonging to same category appear together in the order of questionnaire in the forms. The columns of $\mathcal{T}_{ik}$ correspond to the set of variables $\{a_{ij}\}$ belonging to category $c_{ik}$. Using this procedure across all survey years and categories, we generated a total of 2,056 tables.

%% file: src/lostbench.tex
\section{LoStBench --- Case Study}
\label{sec:case-study}

When selecting a case study for our experiments, we wanted a long-running longitudinal study with publicly available data and a well-curated ground-truth schema.
The American National Election Studies (ANES) Survey~\cite{anes2026_codebook} is a long-running program of academically administered national surveys conducted before and after U.S. presidential elections, and fits these criteria precisely.
ANES captures a continuous time series of data dating back to 1948, and is widely regarded as a gold standard in political science research.
Questions, answers, and metadata in ANES are all representative of the use cases outlined in \Cref{sec:inpractice}, but it is distinct, and so better allows us to assert the generality of our results.
Crucially, the ANES dataset is released alongside a well documented, hand-curated ground-truth target schema that explicitly declares how each source attribute maps into it\footnote{
	Although not all elements of ANES are released in machine readable form, our artifacts submission includes scripts required to extract this data from codebooks.
}.
The presence of an exact ground truth makes ANES an ideal candidate for LoStBench.

\subsection{The ANES Dataset}
\label{lostbench_tables_synthesis}

The ANES dataset contains 32 biennial time-series studies spanning the years 1948 to 2024, and instance-level data summaries are released alongside metadata, data dictionaries, and other documentation\footnote{We note that the project's metadata standards have likewise evolved over its lifetime, and present a data integration challenge of their own.}.
Following the notation introduced in \Cref{sec:preliminaries}, the survey explicitly provides (i) a ground truth schema $\targetschema$, (ii) source datasets with schemas $\schema_1, \ldots, \schema_{32}$, and (iii) mappings $\mapfn_i : \schema_i \rightarrow \targetschema$.
Each dataset consists of responses to questions (attributes) $\attr \in \schema_i$, and for each question ANES provides several forms of metadata that can be used for attribute alignment.
For each of the attributes (i.e., $\left|\targetschema \cup (\bigcup_i \schema_i)\right|$), we obtained:
\begin{itemize}[leftmargin=0.0in,label=]
\item $\attrprop{label}(\attr)$ --- a short, mnemonic, semantically meaningful identifier;
\item $\attrprop{prose}(\attr)$ --- the prose text of the question asked;
\item $\attrprop{dom}(\attr)$ --- the domain of responses to the question or their encodings (e.g. 1--5 for likert scale responses);
\item $\attrprop{category}(\attr)$ --- a thematic group of questions to which the question belongs.
Notably, categories vary across years.
\end{itemize}
We emphasize that these classes of metadata are representative of the types of metadata found across all of our motivating use cases.

Attribute mappings and several forms of metadata were not available in machine-readable form, but rather presented in codebooks.
As the codebooks for each dataset followed largely consistent formatting patterns, and so we were able to script the data extraction process.
Extracted mappings and metadata were cross-checked and discrepancies manually resolved.
Using this process, we obtained source variable metadata for 30 survey years:
(i) We omitted the 1996 release due to missing category information,
and (ii) We excluded the 2024 release as it was incomplete at the time of our analysis.
We obtained 10,320 source-to-target mappings (i.e., the range of all $\mapfn_i$) across the 30 years.

\subsection{Contrast with Other Benchmark Datasets}

\Cref{tab:sm_benchmarks_comparison} contrasts the ANES dataset with benchmark datasets frequently used to evaluate non-longitudinal attribute alignment systems.
In addition to baseline statistics, we report the Szymkiewicz--Simpson coefficient\footnote{$ssc(A, B) = \frac{|A \cap B|}{\min(|A|, |B|)}$} to quantify overlaps between pairs of tables $(\schema_i,\schema_j)$. ANES has 30 schema versions, so we consider 29 pairs of temporally adjacent versions as table pairs $(\schema_i,\schema_{i+1})$. For attribute names, overlap is computed over the sets $\{\attrprop{label}(\attr) \mid \attr \in \schema_i\}$ and $\{\attrprop{label}(\attr) \mid \attr \in \schema_j\}$, and likewise for data instances using the corresponding domains $\attrprop{dom}(\cdot)$.
A value of 1 indicates identical schemas (resp., data domains), while a value of 0 indicates no overlap; Lower values indicate greater diversity in the dataset.
For value overlap, we ignore numeric-valued columns.

The number of attributes (i.e., questions) in ANES is significantly higher than other benchmarks, with even the smallest schema over four times the median table size in the other benchmarks, and the largest almost three times as large.
Name overlap is low, in fact among the lowest in the lot. Value overlap, conversely, is moderately higher (around the median across benchmarks) due to question patterns with standardized response categories such as ``YES'', ``NO'', ``DON'T KNOW'', or ``AGREE''.

\begin{table}[h]
\centering
\scriptsize
\setlength{\tabcolsep}{1pt}
\begin{tabular}{lcccccccc}
\toprule
Dataset & \#Table Pairs & \#Cols & \#Rows & Match Type 
& \multicolumn{2}{c}{Name Overlap} 
& \multicolumn{2}{c}{Value Overlap} \\
\cmidrule(lr){6-7} \cmidrule(lr){8-9}
& & & & 
& Mean & Variance 
& Mean & Variance \\
\midrule

GDC & 10 & 16--736 & 93--4.5k & Human-Curated 
& \colorbox{green!60}{0.049} & 0.009
& \colorbox{green!60}{0.077} & 0.004\\

Magellan & 7 & 3--7 & 0.9k--131k & Human-Curated 
& \colorbox{red!58}{1.0} & 0.0
& 0.315 & 0.039\\

WikiData & 4 & 13--20 & 5.4k--10.8k & Human-Curated 
& 0.076 & 0.003
& \colorbox{red!58}{0.581} & 0.097\\

Open Data & 179 & 26--51 & 11.6k--23k & Fabricated 
& 0.095 & 0.063
& 0.309 & 0.078\\

ChEMBL & 180 & 12--23 & 7.5k--15k & Fabricated 
& 0.10 & 0.067
& 0.148 & 0.102\\

TPC-DI & 176 & 11--22 & 7.5k--15k & Fabricated 
& 0.104 & 0.070
& 0.211 & 0.104\\

\textbf{ANES} & 29 & 67-2.2k & 0.6k--8.3k & Human-Curated 
& \colorbox{green!30}{0.074} & 0.011
& \colorbox{red!9}{0.255} & 0.011\\

\bottomrule
\end{tabular}

\caption{Comparative Summary of ANES schemas with other popular schema matching benchmark datasets}
\label{tab:sm_benchmarks_comparison}
\trimfigurespacing
\vspace*{-2mm}
\end{table}

%% file: src/goals.tex
\section{LoStBench --- Methodology}
\label{sec:long-data-int-goal}

\newtheorem{question}{RQ}

Our primary goal in this paper is to understand the degree to which existing techniques for schema matching and adjacent problems can be re-used for longitudinal data integration tasks.
By extension, in situations where they are not sufficient, we would like to understand why, and to propose fixes; Indeed, we have already done so in \Cref{sec:wide-table}.
This give us our overarching research question, which we address through three more targeted questions.

\begin{question}
  \label{q:effectiveness}
  How effective are existing attribute alignment schemes on longitudinal data integration tasks?
\end{question}

%%%%%%%%%%%%%%%%%%%%%%%%%%%%%%%%%%%%%%%%%%%%%%%%%%%%%%%%%%%%%%%%%%%%%%%%%%%%%%%%%%%%%%%%%%%%%%%%
%%%%%%%%%%%%%%%%%%%%%%%%%%%%%%%%%%%%%%%%%%%%%%%%%%%%%%%%%%%%%%%%%%%%%%%%%%%%%%%%%%%%%%%%%%%%%%%%
%%%%%%%%%%%%%%%%%%%%%%%%%%%%%%%%%%%%%%%%%%%%%%%%%%%%%%%%%%%%%%%%%%%%%%%%%%%%%%%%%%%%%%%%%%%%%%%%

\subsection{Pairwise Alignment}
First, we isolate differences that arise from the nature of the data itself: We want to know how effective the existing schema matching algorithms are at working with ANES schemas and data.
As we already noted, existing alignment schemes struggle with the wide schemas of ANES, but there are a range of other factors that we want to consider.
For example, the prose questions ($\attrprop{prose}(\cdot)$) carry a much stronger signal about alignment than attribute labels ($\attrprop{label}(\cdot)$), but we want to know whether existing schemes are capable of using this signal.

\begin{question}
  \label{q:pairwise}
  How effective are existing attribute alignment schemes when used to align pairs of survey revisions?
\end{question}

We address this question by evaluating each form of attribute alignment on its ability to align pairs of schemas $\schema_i$ and $\schema_j$ drawn from the ANES data.
The alignment technique's success is evaluated through three measures: (i) Recall@GroundTruth (R@GT), (ii) Mean Reciprocal Rank (MRR), and (iii) Rank For Ideal Recall (RIR).
The first measure models the technique in an idealized setting, measuring its ability to act without supervision.
An ideal ranking of $\text{R@GT} = 1$ suggests that the algorithm may be used with minimal supervision.

\begin{definition}[\textbf{R@GT}]
\label{def:ragt}
Let $\gtmatching$ be the ground-truth matching for a pair of schemas $\schema_i$, $\schema_j$.
We are given a top-$\abs{\gtmatching}$ matching $\matching$ (\Cref{def:mrank});
We define the Recall@GroundTruth for $\matching$ as the fraction of it that is correct:
\[
\ragt = \frac{\abs{\matching \cap \gtmatching}}{\abs{\gtmatching}}
\]

\end{definition}

Unlike R@GT, which considers a global ranking of all candidate matches across variables, a \emph{mean reciprocal rank} (MRR) evaluates the rankings individually for each source variable.
It is a standard metric commonly used in evaluating ranked retrieval systems, such as search engines and question-answering systems~\cite{IR_textbook}.

\begin{definition}[\textbf{MRR}]
\label{def:mrr}
Let $\gtmatching$ be the ground-truth matching for schemas $\schema_i$, $\schema_j$ and recall that $\arank{\attr}{\attr'}$ is the rank of $\attr'$ with respect to $\attr$.
Define the score of $\attr \in \schema_i$ to be the minimal rank of a correct attribute matching for $a$:
$$\textsc{score}(\attr) = \min_{\intuple{\attr, \attr'} \in \gtmatching} \arank{\attr}{\attr'}$$ 
Let $\mu$ be the set of attributes with at least one correct match:
$\mu = \comprehension{\attr}{\intuple{\attr, \_} \in \gtmatching}$.
The mean reciprocal rank is the average of the reciprocal ranks for all matching attributes:

\[
\mrr = \frac{1}{|\mu|} \sum_{\attr \in \mu} \frac{1}{\textsc{score}(\attr)}
\]
\end{definition}

Intuitively, MRR captures how quickly the first relevant match appears when examining the ranked list of candidate matches. A higher MRR indicates that the algorithm ranks correct matches near the top of the list, allowing users to more easily identify them in human-in-the-loop settings.
But MRR is only informative and helpful until we are looking for 1 relevant match.
If we are matching more than two schemas, or a variable has multiple matches, then in such cases MRR doesn't quantify how far along we need to look for in a ranked list for all relevant matches.
The further down the list the correct match appears, the more work the manual curator must perform.
To quantify this, we propose a new measure: \textbf{Rank For Ideal Recall}.

\begin{definition}[\textbf{RIR}]
\label{max_rank_4_ideal_recall}
Let $\gtmatching$ be the ground-truth matching for schemas $\schema_i$, $\schema_j$ and recall that $\arank{\attr}{\attr'}$ is the rank of $\attr'$ with respect to $\attr$.
Given $\attr \in \schema_i$, the rank for ideal recall of $\attr$ in $\gtmatching$ is:
\[
\rir(\attr) = \max_{\intuple{\attr, \attr'} \in \gtmatching} \arank{\attr}{\attr'} 
\]
The RIR for an alignment is the sum over matched attributes:
$\rir = \sum_{\attr \in \mu} \rir(\attr)$
where $\mu$ is defined as in Definition~\ref{def:mrr}.
\end{definition}

%%%%%%%%%%%%%%%%%%%%%%%%%%%%%%%%%%%%%%%%%%%%%%%%%%%%%%%%%%%%%%%%%%%%%%%%%%%%%%%%%%%%%%%%%%%%%%%%
%%%%%%%%%%%%%%%%%%%%%%%%%%%%%%%%%%%%%%%%%%%%%%%%%%%%%%%%%%%%%%%%%%%%%%%%%%%%%%%%%%%%%%%%%%%%%%%%
%%%%%%%%%%%%%%%%%%%%%%%%%%%%%%%%%%%%%%%%%%%%%%%%%%%%%%%%%%%%%%%%%%%%%%%%%%%%%%%%%%%%%%%%%%%%%%%%

\subsection{Category Alignment}
As outlined in \Cref{sec:wide-table}, some existing alignment techniques lack support for ANES's wide schemas.
To address this limitation, we leverage the hierarchical structure of survey instruments like ANES to restrict schema matching to within matched categories.
However, this optimization is predicated on the attribute alignment technique being able to correctly match categories.
We next ask whether this is the case.

\begin{question}
\label{q:category}
  Can attribute alignment schemes effectively match categories across schema revisions?
\end{question}

We answer this question by re-interpreting LoStBench as a table unionability benchmark, partitioning schemas into individual per-category (i.e., $\attrprop{category}(\cdot)$) tables, and drawing a sample as described in \Cref{sec:experiments}.
\Cref{tab:tus_benchmarks} contrasts LoStBench with other table unionability benchmarks.
\citet{boutaleb-etal-2025-somethings} identified three key limitations in such benchmarks: semantic simplicity, ground-truth inconsistencies, and excessive overlap.
As previously noted, semantic simplicity is a limitation of the longitudinal study domain.
However, the survey questions in ANES cover a wide range of general social science variables, including personal, economic, educational, and political topics, providing meaningful diversity for analysis.
Ground-truth inconsistencies are expected to be minimal, as the dataset is derived from a source with over 10,000 citations, maintained manually for more than 75 years, and periodically corrected. In terms of overlaps, LoStBench exhibits the lowest name overlap and the second-lowest value overlap among the benchmarks.

\begin{table*}[t]
\centering
\footnotesize
% \scriptsize
\setlength{\tabcolsep}{2pt}
\begin{tabular}{llrrrrrrrrrr}
\toprule
Benchmark & Type & Table Category &\#Tables & Avg Shape & Null(\%) & Numeric(\%) & Non-Numeric(\%)& Size (MB) & Name Overlap & Value Overlap \\
\midrule
\multirow{2}{*}{UGEN V2}
& \multirow{2}{*}{LLM-Generated} & Candidates & 1000  & 19 $\times$ 14 & 5.87 & 17.17 & 82.83  & \multirow{2}{*}{$\sim$7} &\multirow{2}{*}{0.257} &\multirow{2}{*}{0.422}\\
& & Query  & 50  & 108 $\times$ 14 & 4.45 & 12.33 & 87.67 & \\

\midrule

\multirow{2}{*}{SANTOS}
& \multirow{2}{*}{Fabricated}& Candidates & 500  & 6922 $\times$ 12 & 7.61 & 25.64 & 74.36  & \multirow{2}{*}{$\sim$440} &\multirow{2}{*}{0.781} & \multirow{2}{*}{0.674}\\
& & Query  & 50  & 21402 $\times$ 13 & 5.88 & 23.90 & 76.1 & \\

\midrule

\multirow{2}{*}{Real Tables}
& \multirow{2}{*}{Fabricated} & Candidates & 11073  & 7719 $\times$ 11 & 22.60 & 26.03 & 73.97 & \multirow{2}{*}{$\sim$11,500} & \multirow{2}{*}{0.837} &\multirow{2}{*}{0.601}\\
& & Query  & 80  & 12918 $\times$ 13 & 16.90 & 26.35 & 73.65 & \\

\midrule

\multirow{2}{*}{TUS Large}
& \multirow{2}{*}{Fabricated} & Candidates & 5043  & 1915 $\times$ 11 & 12.53 & 8.67 & 91.33 & \multirow{2}{*}{$\sim$1,590} &\multirow{2}{*}{0.680} &\multirow{2}{*}{0.509}\\
& & Query  & 4296  & 2132 $\times$ 18 & 14.87 & 7.81 & 92.19 & \\

\midrule
\multirow{2}{*}{Lakebench-OpenData}
&Fabricated \& & Candidates & 4832  & 112,360 $\times$ 21 & 3.44 & 44.93 & 55.07 &  \multirow{2}{*}{\colorbox{green!30} {$\sim$80,000}} &\multirow{2}{*}{0.613} &\multirow{2}{*}{0.453}\\
& Human-Curated & Query  & 3138  & 79,311 $\times$ 17 & 2.90 & 53.87 & 46.13 & \\

\midrule

\multirow{2}{*}{PYLON}
& \multirow{2}{*}{Human-Curated} & Candidates & 1746  & 115 $\times$ 10 & - & 41.26 & 58.74 & \multirow{2}{*}{$\sim$22} &\multirow{2}{*}{0.155} &\multirow{2}{*}{\colorbox{green!30}{0.252}}\\
& & Query  & 124  & 90 $\times$ 7 & - & 24.32 & 75.68 & \\

\midrule

\multirow{2}{*}{\textbf{ANES/LoStBench}}
& \multirow{2}{*}{Human-Curated}& Candidates & 1386  & 2417 $\times$ 9 & \colorbox{red!30}{32.05} & 3.55 & \colorbox{green!30}{96.45} & \multirow{2}{*}{$\sim$385} &  \multirow{2}{*}{\colorbox{green!30}{0.088}} & \multirow{2}{*}{\colorbox{green!10}{0.281}}\\
& & Query  & 115  & 2049 $\times$ 9 & \colorbox{red!30}{28.36} & 4.40 & \colorbox{green!30}{95.60} & \\

\bottomrule
\end{tabular}

\caption{Table Union Search Benchmarks Comparison}
\label{tab:tus_benchmarks}
\trimfigurespacing
\end{table*}

% Measures for ground-truth inconsistencies such as ground-truth false positive/negative rates (GTFP@K and GTFN@K \cite{thakur2021beir}) indicate the fraction of top-k candidates retrieved by a search method that are not labeled as required in the original ground truth.
% These measures rely on the reliability of the model generating the ranked list and may overestimate errors in cases where a candidate's exclusion from the ground truth is actually correct because of nuances in table definitions or descriptions that the model could not identify but only human could.
The LoStBench table unionability workload is comparable in shape and cardinality to other table unionability workloads, but differs significantly in terms of its data.
Like many social science surveys, participants may elect to omit answers and almost a third of the data is null.
By contrast, even the most sparse table unionability benchmark dataset has only a quarter null values.
We also note that in contrast to other table unionability benchmarks, where null values indicate the absence of semantic content, an individual's choice to skip an answer is often semantically meaningful to statisticians using datasets such as ANES.
A second distinguishing feature of ANES is that responses are frequently categorical, and rarely (<5\%) numerical in contrast to other benchmark datasets.
We note that comparatively few table unionability schemes support numerical data, so this is a positive feature for existing approaches.

When evaluating table unionability, we focus on three measures. 
% These measures evaluate how well do existing alignment techniques retrieve top-$k$ unionable schema as defined below:

% \begin{definition}[Top-$k$ Unionable Schema Set]
% Given a query schema $\schema^Q$, the top-$k$ unionable schema set denoted by $\schema'$, is defined as a subset $\schema' \subseteq \bigcup_i \schema_i$ such that $\abs{\schema'}=k$ and:
% % \[
% %     \forall \schema'' \in \schema',\ \forall \schema''' \in \bigcup_i \schema_i \setminus \schema':\ \ 
% %     \similarity(\schema^Q, \schema'') \geq \similarity(\schema^Q, \schema''')
% % \]
% \[
%     \forall \schema'' \in \schema',\ \forall \schema''' \notin \schema':\ \ 
%     \similarity(\schema^Q, \schema'') \geq \similarity(\schema^Q, \schema''')
% \]
% \end{definition}

% \ok{these need to be updated}
\begin{definition}[\textbf{P@K}]
    Precision at rank $K$ measures the fraction of relevant items among the top-$k$ retrieved results.
\begin{equation*}
\text{Precision@}K = \frac{1}{K} \sum_{k=1}^{K} \mathrm{rel}(k)
\end{equation*}
where rel($k$) $\in$ \{0, 1\} indicates whether the item at rank $k$ is relevant (i.e, in the ground truth).
\end{definition}

\begin{definition}[\textbf{AP@K}]
    Average Precision at $K$ computes the average of Precision@k values at positions where relevant items occur, thereby rewarding methods that retrieve relevant items earlier and penalizing for errors in the early ranks.
    \[
    \text{AP@}K = \frac{1}{\min(N, K)} \sum_{k=1}^{K} \text{Precision@}k \cdot \mathrm{rel}(k)
    \]
    where $N$ is the total number of relevant items.
\end{definition}

\begin{definition}[\textbf{MAP@K}]
    Mean Average Precision at $K$ is the mean of AP@K over all queries.
\[
MAP@K = \frac{1}{|Q|} \sum_{q \in Q} \text{AP@}K(q)
\]
where $Q$ is the set of queries and AP@K(q) is the Average Precision at K for query q. 
\end{definition}

%%%%%%%%%%%%%%%%%%%%%%%%%%%%%%%%%%%%%%%%%%%%%%%%%%%%%%%%%%%%%%%%%%%%%%%%%%%%%%%%%%%%%%%%%%%%%%%%
%%%%%%%%%%%%%%%%%%%%%%%%%%%%%%%%%%%%%%%%%%%%%%%%%%%%%%%%%%%%%%%%%%%%%%%%%%%%%%%%%%%%%%%%%%%%%%%%
%%%%%%%%%%%%%%%%%%%%%%%%%%%%%%%%%%%%%%%%%%%%%%%%%%%%%%%%%%%%%%%%%%%%%%%%%%%%%%%%%%%%%%%%%%%%%%%%

\subsection{Longitudinal Alignment}

In \Cref{sec:attribute-alignment}, we propose a strategy for generalizing pairwise schema matching to matching a sequence of schemas.
This approach comes with a trade-off: More steps are required, but each step is significantly more targeted --- the curator has fewer potential alignments to review (and discard) in each step.
Our final question considers the utility of this approach.

\begin{question}
\label{q:incremental-alignment}
    Does incremental alignment reduce the effort required to integrate a longitudinal study.
\end{question}

Rank for Ideal Recall (RIR)  (Definition \ref{max_rank_4_ideal_recall}) measures the number of matchings that the user must consider in order to fully integrate a dataset, and so we use this measure as a proxy for effort.

%% file: src/experiments.tex
\section{Experiments}
\label{sec:experiments}

Our overarching goal (RQ~\ref{q:effectiveness}) is to understand how effective existing schema matching algorithms can be when applied to the iterative evolution of a longitudinal study.
We now address this goal through experiments that
(i) use pairwise schema matching to isolate domain-specific effects of ANES on schema matching algorithms (RQ~\ref{q:pairwise}, addressed in \Cref{sec:schema-matching-experiment});
(ii) confirm the value of incremental matching (RQ~\ref{q:incremental-alignment}, addressed in \Cref{sec:multiple-merging-schemas-experiment}); and
(iii) examine the efficacy of the blocking optimization from \Cref{sec:wide-table} (RQ~\ref{q:category}, addressed in \Cref{sec:table-union-experiment}).

\paragraph{Experimental Setup}
\Cref{sec:preliminaries} reviews and motivates the choice of systems being evaluated, and section \ref{q:incremental-alignment} documents how we adapted these systems to the setting of longitudinal data integration, and to wide attribute schemas.
Most schema matching systems are intended for use with attribute identifiers $\attrprop{label}(\attr)$, but because of the additional semantic value present in $\attrprop{prose}(\attr)$, we attempted to use the latter as an identifier with each system.
We report these results if they were better.

For COMA/COMA++, Cupid, Sim Flooding, Distr-based, and Jaccard-Lev, we used the publicly available implementation from \textsc{Valentine}~\cite{valentine}.
We ran these algorithms with all input parameter combinations suggested in the original paper, and report results from the optimal combination of settings.
The results of Cupid \cite{cupid} were significantly worse than other approaches, and so are omitted from our reporting.
For \textsc{Magneto} we used it's publicly available implementation, whereas \textsc{DDM} is our own previous work.
The embedding model used for DDM was \emph{paraphrase-MiniLM-L6-V2}.
We evaluated Magneto using its zero-shot configurations (\emph{Magneto-zs-bp} and \emph{Magneto-zs-llm}) and did not perform fine-tuning.
In the general schema alignment case, we do not expect labeled data for fine-tuning to be readily available, requiring practitioners to rely on off-the-shelf models.
Furthermore, attribute semantics are largely conveyed through general-purpose natural language in question text, which aligns well with the pretraining distribution of modern language models like \emph{mpnet} used by Magneto.
We also experimented with each of Magneto's serialization strategies (default, verbose, and repeat), and got slightly better results for verbose serialization compared to the others; we report these results only.
For \emph{MagnetoGPT} and \emph{DDM\_LLM} we used \emph{gpt-4o-mini} as LLM re-ranker. 

We used the publicly available implementations of SANTOS, Starmie, D3l, and MetaTUS.
In the case of \textsc{SANTOS} experiment, it was additionally necessary to identify \textit{intent columns}, a process they performed manually in their own evaluation.
In LoStBench, we generated intent columns by using the attribute with the highest number of matched attributes in the ground truth.
We pretrained \textsc{Starmie} on LoStBench by exploring a subset of the combinations of hyperparameter space, varying augmentation operators and sampling methods while keeping other parameters at their recommended values as mentioned in the documentation.
Then we performed the search with \texttt{naive}, \texttt{HNSW}, and \texttt{LSH} techniques on different threshold and k-values on recommended hyperparameters.

The preprocessing and index generation for \textsc{SANTOS} were carried out on a machine with 1~TB of memory to accommodate the in memory processing of raw DBpedia knowledge base, consisting of single files of over 100~GB each.
All other workloads for the methods described above were executed on a machine running \textsc{Pop!}\_OS~22.04~LTS (Ubuntu~22.04), equipped with an AMD Ryzen~7~8700G processor with 8 physical cores and 16 threads, and 64~GB of system memory.
GPU-based methods utilized the GPU for computation, except for the scalability experiment, which was run on the CPU to maintain consistency with the other methods.

%%%%%%%%%%%%%%%%%%%%%%%%%%%%%%%%%%%%%%%%%%%%%%%%%%%%%%%%%%%%%%%%%%%%%%%%%%%%%%%%%%%%%%
%%%%%%%%%%%%%%%%%%%%%%%%%%%%%%%%%%%%%%%%%%%%%%%%%%%%%%%%%%%%%%%%%%%%%%%%%%%%%%%%%%%%%%
%%%%%%%%%%%%%%%%%%%%%%%%%%%%%%%%%%%%%%%%%%%%%%%%%%%%%%%%%%%%%%%%%%%%%%%%%%%%%%%%%%%%%%
\subsection{Experiment 1: Schema Matching}
\label{sec:schema-matching-experiment}

Our first experiment measures the effectiveness of existing schema matching algorithms on pairwise matching.
Recall that ANES attributes are organized into groups of related questions called categories.
We model the wide-schema optimization from \Cref{sec:wide-table}, by selecting matched pairs of categories from adjacent survey iterations and evaluating the efficacy of the schema matching tool on the pair.
For constructing the test set for schema matching, we aimed for a reasonable pair count because we were dealing with methods making api calls to LLMs, that could potentially take a minute to run for each pair.
Additionally, we aimed to maximize schema diversity.
To this end, we ensured that no category appeared more than twice across the sampled test cases.
We performed stratified sampling based on the number of linked variables in each category pair.
Since the distribution was skewed, most pairs had only two linked variables with progressively fewer pairs at higher link counts.
With this setup, preliminary experiments resulted in the discovery of two significant limitations of existing schema matching systems that we discuss below.

\paragraph{Missing Values}
Consider the following block of questions:
\begin{quote}
\footnotesize
\textit{
\textbf{D5.} (OCCUPATION OF HEAD OF HOUSEHOLD) WHAT IS (YOUR/THE HEAD OF THE HOUSEHOLD'S) OCCUPATION? I MEAN, WHAT KIND OF WORK (DO YOU/DOES HE) DO? \\
\textbf{D6.} (DO YOU/DOES HE) DO ANY OTHER KIND OF WORK BESIDES FARMING--FOR PAY, I MEAN?\\
THE FOLLOWING THREE QUESTIONS ARE FOR FARMERS ONLY:\\ 
\textbf{D7.} (DO YOU/DOES HE) OWN (YOUR/HIS) FARM, MANAGE IT, RENT IT, OR WHAT?\\
\textbf{D8.} HOW MUCH LAND (DO YOU/DOES HE) FARM?\\
\textbf{D9.} WHAT KIND OF FARMING (DO YOU/DOES HE) DO HERE MOSTLY?
}
\end{quote}
In years where no farmers participated in a given survey, response to the questions (D7-D9) are entirely empty.
This happens quite frequently --- of the 2,056 categories seen in 30 iterations of ANES datasets, 670 have at least one column completely empty.
Cases like these convey no semantic content for instance-based models, making them effectively useless.
For further experiments, we restricted ourselves to the remaining 1,386 categories with non-empty instances.

\begin{figure}[!t]
    \centering
    \includegraphics[width=1.01\linewidth]{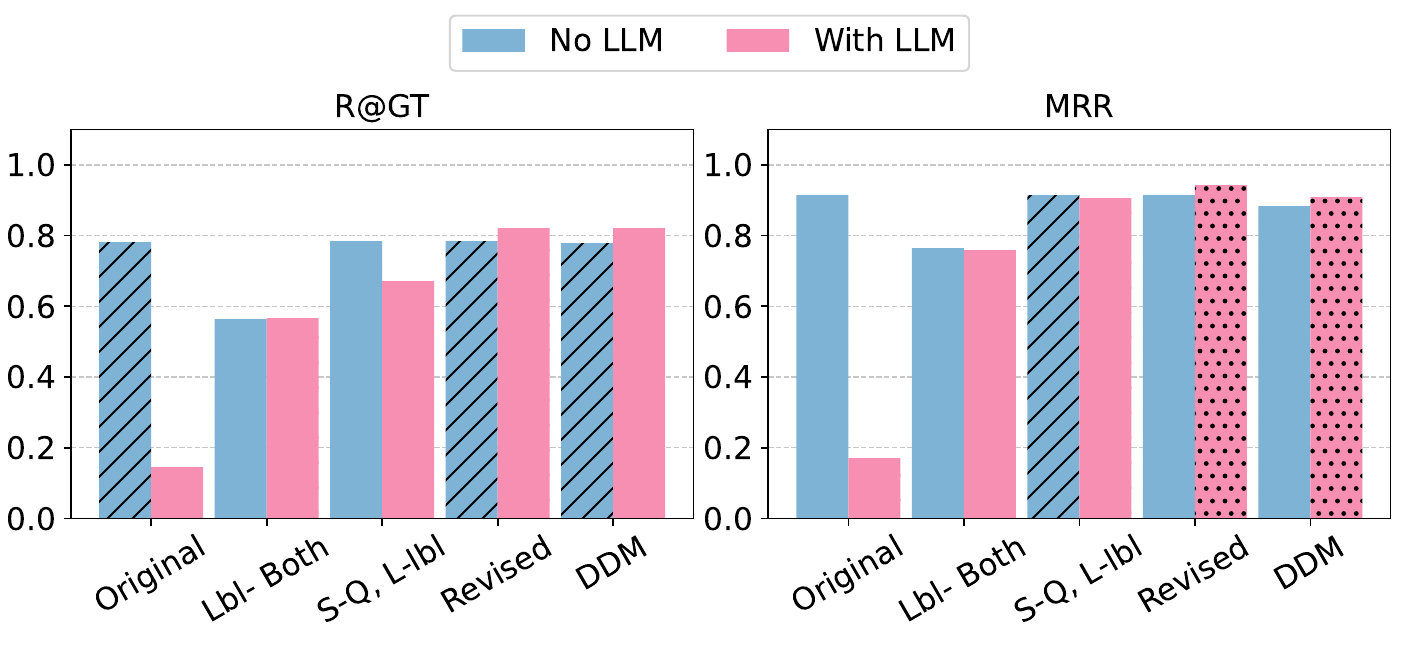}
    \caption{Improvement of LLM re-ranker with changes in prompts. (Note: S-Q, L-lbl means SLM-Qn, LLM-lbl)}
    \label{fig:llm_vs_no_llm}
    \trimfigurespacing
\end{figure}

\begin{figure*}[t]
    \centering
    
    % Top subfigure
    \begin{subfigure}[t]{0.78\textwidth}
        \centering
        \includegraphics[width=\textwidth]{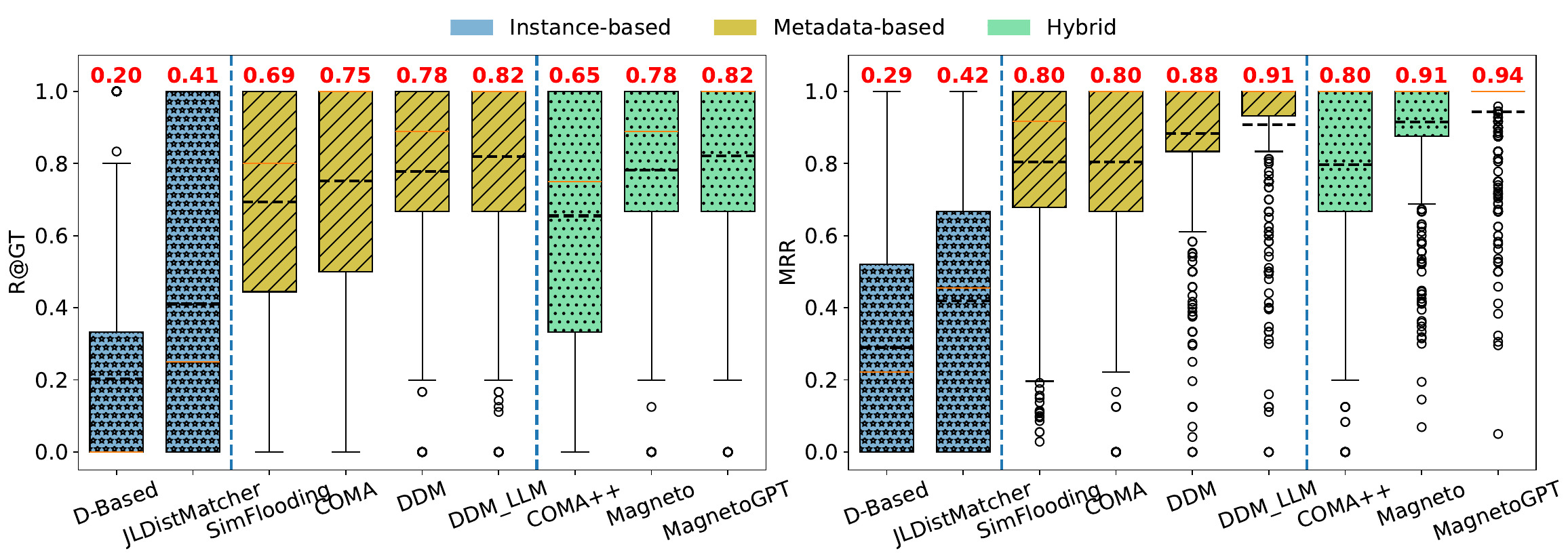}
        \caption{Recall@GT and MRR. Red annotations are Average values.}
        \label{fig:recall_at_gt_compare_sub}
    \end{subfigure}
    
    \vspace{0.01cm}
    
    % Bottom subfigure
    \begin{subfigure}[t]{\textwidth}
        \centering
        \includegraphics[width=0.78\textwidth]{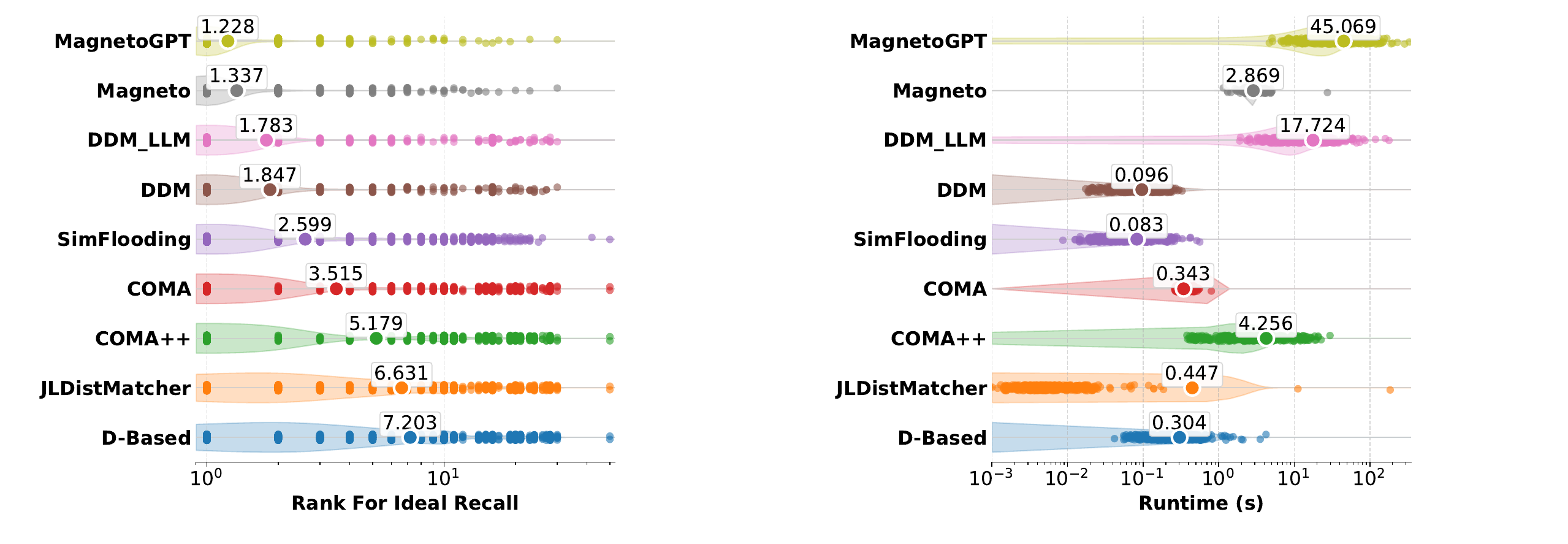}
        \caption{Ranks for Ideal Recall and Runtimes. Lower is Better.}
        \label{fig:max_ranks_4_ideal_recall_sub}
    \end{subfigure}

    \caption{Comparison of Schema Matching Methods}
    \label{fig:combined_models_comparison}
    \trimfigurespacing
\end{figure*}

\paragraph{Redundancy}
Redundancy also plays a significant role in survey design, with repeated or rephrased questions used to detect bias or correct errors~\cite{DBLP:conf/ictd/ChenCCDHP09}.
Many schema matching and table unionability algorithms rely on heuristics that reduce the search space by explicitly excluding duplicate matches (e.g., c-alignment~\cite{3.fatemeh}).
Such heuristics are not appropriate for survey data.
For further experiments, we eliminated many-to-many and one-to-many attribute pairings.

Considering the missing values and redundancies, we refined our sampling process to omit parts of the problem domain that existing systems are incapable of addressing.
From this point we applied weighted sampling- more weights to pairs with higher no. of links- to downsample to 600 category pairs.
To further balance the distribution, we randomly removed 150 pairs containing only two linked variables, resulting in a final set of 450 category pairs for the test set, twice the size of any other benchmark that appears in table~\ref{tab:sm_benchmarks_comparison}.

\paragraph{LLMs and Question Identifiers}
\label{llm_inability}

Magneto leverages both variable names and column values for its SLM and LLM components to infer column semantics, and as noted above, we use the question prose ($\attrprop{prose}(\attr)$) in lieu of the attribute name where it is beneficial to do so.
We identified two main issues with their original llm prompt layout when using $\attrprop{prose}(\attr)$ as the attribute identifier.
While the results for \emph{Magneto-zs-bp} were good, the performance of the Magneto LLM re-ranker was nearly five times worse than the SLM component.
For the original LLM prompt employed in \emph{Magneto-zs-llm}, the LLM frequently failed to produce output that could be successfully parsed, and in several cases the parser had to abort for a variable after five retries.  
One problem was that the LLM was unable to consistently reproduce the long prose attribute names.
For instance, one question was titled ``Q.J6. What about the next 12 months?Do you...''
When identifying to this question in the re-ranking step, the LLM might inject an additional space between the '?' and the 'D'.
Though interpretable to a human, this alteration disrupted Magneto's parser.
In other cases, the same attribute captures the responses to multiple linked sub-questions under a single merged scale (e.g. ``Q.J5a.'' and ``Q.J5b.'').
Because the prose for these questions was concatenated, the LLM's output would treat each sub-question as an independent question, leading to more parser disruption.

\Cref{fig:llm_vs_no_llm} illustrates a representative schema matching performed using a variety of systems, both in the raw form (No LLM) and with an LLM based re-ranking step.
Notably, using $\attrprop{prose}(\attr)$ in place of attribute name, the LLM re-ranking step decreased accuracy, resulting in an accuracy almost five times worse.
Reverting to $\attrprop{label}(\attr)$ as an attribute name (Lbl- Both) prevented LLM parser failures, and in turn prevented re-ranking errors.
However, this came at the cost of a substantial reduction in accuracy of the pre-LLM step, leading to an overall reduction in accuracy.
We also explored a third, hybrid approach (S-Q, L-lbl) by modifying Magneto to use $\attrprop{prose}(\attr)$ in the initial SLM step, and $\attrprop{label}(\attr)$ as a variable identifier in the LLM prompt.
This preserved the SLM component’s performance but with the absent the prose signal, the re-ranking step reduces accuracy.

Our final effort generates a re-ranking prompt that uses $\attrprop{label}(\attr)$ as an identifier, but also provides $\attrprop{prose}(\attr)$ in a ``Detailed meaning of the column'' field within the prompt.
This revision resulted in a slight improvement in accuracy from re-ranking.
We explored a similar revision in the context of DDM; Leveraging its insight that context plays a role in longitudinal studies, we also provided the LLM with all the preceding variables in the schema when prompting the LLM for a given variable in the schema.

\paragraph{Comparative Analysis}
\Cref{fig:recall_at_gt_compare_sub} shows pairwise schema matching performance.
The results show a clear performance stratification across the evaluated schema matching algorithms.
Even with missing values omitted, instance-based methods perform poorly.
ANES questions follow a set of common patterns, and so relying solely on literal data values provides limited semantic information to distinguish between differences in column meanings.
In contrast, metadata-based and hybrid approaches benefit from the richer semantic content present in the associated prose questions.

% As we discussed in Section~\ref{schema_matching}, since \emph{Recall@GT} and \emph{MRR} alone may not fully capture the overhead imposed on human curators who must validate and correct the ranked lists of match suggestions. To better assess this overhead, we consider the \emph{Rank for Ideal Recall}, as defined in Definition~\ref{max_rank_4_ideal_recall}. 
Figure~\ref{fig:max_ranks_4_ideal_recall_sub} shows that the top four models in terms of \emph{Recall@GT} and \emph{MRR} also exhibit the lowest RIR values, albeit at the cost of substantially increased runtime.
Recall that RIR serves as a proxy for human-in-the-loop effort: Magneto's 1.34 RIR indicates that in expectation, all correct matchings for an attribute appears within the first 1.34 predicted pairings for an attribute.
By contrast, DDM, a simple approach based on word embeddings, is also able to place the correct result within the top 2 positions on average, while executing in only a tenth of a second compared to nearly 3 seconds for Magneto, or three quarters of a minute for MagnetoGPT.

\paragraph{Ablation Study: Frequency of LLM Hallucination and Re-Ranker Upper Bound}

While the LLM re-ranker improved results in several cases, we observed that it often underperformed compared to the SLM component whose ranked output it consumes.
To better understand this behavior, we conducted an ablation study to quantify how frequently the LLM re-ranker produces stochastic errors (hallucinations) relative to its SLM input and to estimate an upper bound on Recall@GT.
We performed this experiment using DDM.
For each test instance, we compared the Recall@GT of the LLM re-ranked output with that of the original SLM-ranked list.
Whenever the LLM re-ranker produced a lower Recall@GT than the SLM, we provided feedback and prompted the LLM to regenerate the ranked list using the same input.
This process was repeated until the LLM achieved a Recall@GT at least equal to that of the SLM, or until a maximum of 10 retries was reached, after which the instance was skipped.
This setup allowed us to estimate both the frequency of stochastic errors and the achievable upper bound under iterative correction.
We observed that in \emph{90} out of \emph{450} instances (20\% instances), the LLM re-ranker required at least one retry, indicating a stochastic error.
Under this retry-based regime, the average Recall@GT improved to $0.827$, compared to $0.821$ in the standard single-pass setting (as shown in Figure~\ref{fig:llm_vs_no_llm}).
These results suggest that while the LLM re-ranker has the potential to improve performance, it is prone to instability, and its gains are bounded even under idealized corrective feedback.

%%%%%%%%%%%%%%%%%%%%%%%%%%%%%%%%%%%%%%%%%%%%%%%%%%%%%%%%%%%%%%%%%%%%%%%%%%%%%%%%%%%%%%
%%%%%%%%%%%%%%%%%%%%%%%%%%%%%%%%%%%%%%%%%%%%%%%%%%%%%%%%%%%%%%%%%%%%%%%%%%%%%%%%%%%%%%
%%%%%%%%%%%%%%%%%%%%%%%%%%%%%%%%%%%%%%%%%%%%%%%%%%%%%%%%%%%%%%%%%%%%%%%%%%%%%%%%%%%%%%
\subsection{Experiment 2: Iterative Schema Matching}
\label{sec:multiple-merging-schemas-experiment}

Our second experiment evaluates the iterative schema matching technique from \Cref{sec:attribute-alignment}.
Concretely, we model a setting where a user reviews and manually corrects the schema matching at each iteration $\matching_i$ before the process proceeds.
We use DDM as a representative schema matching algorithm, as it is the most accurate sub-second runtime algorithm.
Concretely, we use DDM to generate a ranking, compute its $\rir$.
We model the user manually curating the matching by replacing the matching with the correct ground-truth matching.
We consider three variations on this approach:
(i) \Cref{alg:incr-integ} with schemas integrated in temporal order (\textbf{Temporal}),
(ii) \Cref{alg:incr-integ} with schemas integrated in random order (\textbf{Random}), and
(iii) An all-at-once, $N$-way merge where the user instead fully constructs one attribute from the ground-truth schema at a time.
Our objective is to analyze how the merging strategy and schema order influence curator effort, as captured by the AUC (Area Under the Curve) of \emph{RIR}.

\paragraph{Test Set for Iterative Schema Alignment}
Given the set of mappings $\mathcal{H}$, we construct, for each schema $S_i$, a set of schemas $\mathcal{R}_i$ from subsequent years that share correspondences with $S_i$.
\[
\mathcal{R}_i = \{ S_j \mid j > i \ \wedge\ \exists\, (a, a') \in M_{i,j} \text{ such that } a \in S_i \ \wedge\ y \in S_j \}.
\]

We then consider the set $V_i = \mathcal{R}_i \cup S_i$ and define an undirected graph $G_i = (V_i, E)$, where
\[
E = \{ (S_p, S_q) \mid \exists\, (a, a') \in M_{p,q} \text{ such that } a \in S_p \ \wedge\ a' \in S_q \}.
\]

For each $i$, we verify whether $G_i$ is a complete graph, i.e.,
\[
\forall S_p, S_q \in V_i,\; S_p \neq S_q \;\Rightarrow\; (S_p, S_q) \in E.
\]

This process yields 71 such instances of $V_i$ satisfying the completeness condition, which are subsequently used as test cases in the iterative Schema Alignment experiments.

\paragraph{Results}

\begin{figure}[!t]
    \centering
    
    % Top subfigure
    \begin{subfigure}[t]{\columnwidth}
        \centering
        \includegraphics[width=1.01\linewidth]{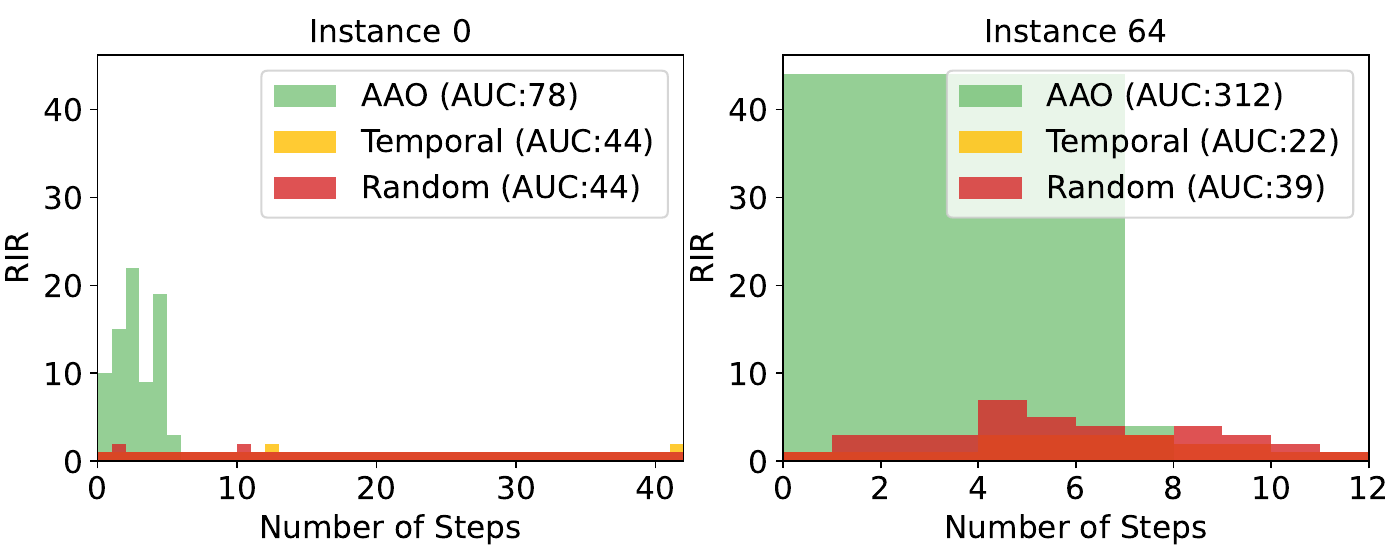}
        \caption{All-at-once (AAO) vs Pairwise Iterative Effort comparison curves in terms of Ranks for Ideal Recall (RIR) for two test case instances when input schema ordering is temporal vs random.}
        \label{fig:aao_vs_pairwise_hitl}
    \end{subfigure}
    
    \vspace{0.01cm} 
    
    % Bottom subfigure
    \begin{subfigure}[t]{\columnwidth}
        \centering
        \includegraphics[width=\linewidth]{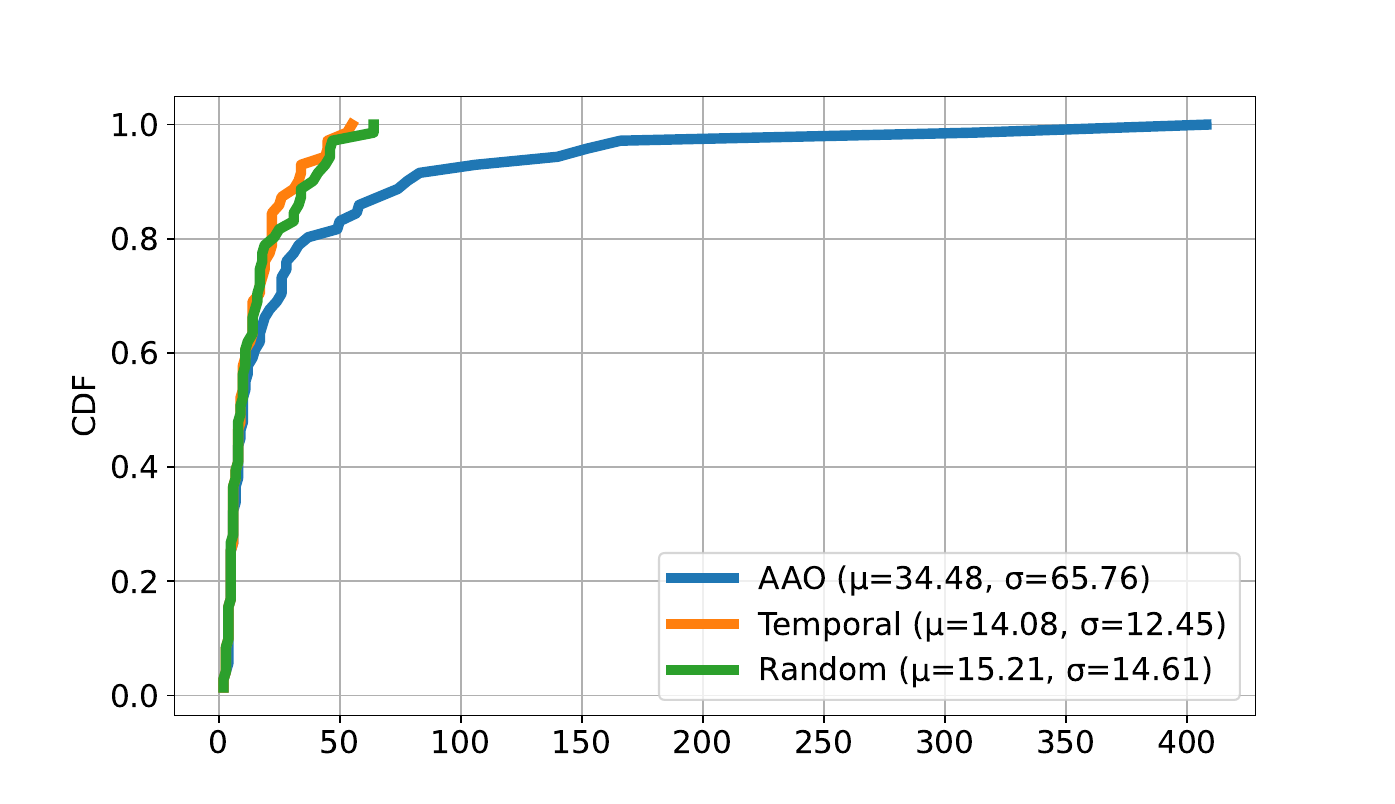}
        \caption{CDF for AUC}
        \label{fig:auc_cdf}
    \end{subfigure}
    
    \caption{Comparison of Merging Techniques}
    \label{fig:merging_techniques_comparison}
    \trimfigurespacing
\end{figure}

Figure~\ref{fig:aao_vs_pairwise_hitl} shows that \emph{all-at-once} alignment yields higher \emph{AUC} than \emph{pairwise iterative} alignment, indicating greater validation and correction effort for the user.
This suggests that curators benefit from distributing effort across smaller, incremental corrections rather than fewer large ones, the answer to our \textbf{RQ \ref{q:incremental-alignment}}.
Notably, the CDF in Figure~\ref{fig:auc_cdf} indicates that a temporal/chronological input ordering of schemas subjected for iterative alignment requires slightly less manual effort than a random input schema ordering when it comes to longitudinal settings.

%%%%%%%%%%%%%%%%%%%%%%%%%%%%%%%%%%%%%%%%%%%%%%%%%%%%%%%%%%%%%%%%%%%%%%%%%%%%%%%%%%%%%%
%%%%%%%%%%%%%%%%%%%%%%%%%%%%%%%%%%%%%%%%%%%%%%%%%%%%%%%%%%%%%%%%%%%%%%%%%%%%%%%%%%%%%%
%%%%%%%%%%%%%%%%%%%%%%%%%%%%%%%%%%%%%%%%%%%%%%%%%%%%%%%%%%%%%%%%%%%%%%%%%%%%%%%%%%%%%%
\subsection{Experiment 3: Table Union Search}
\label{sec:table-union-experiment}

As above, we focus on the 1386 categories on which all approaches have a chance of being effective.
Treating each category as a table, we constructed a ranked list of unionable categories, ordering them by the number of linked variables.
From this set, we stratified and sampled 115 query tables ($\sim$10\%) based on the number of unionable tables associated with each query.

\paragraph{Limitations of Ontology-Based Alignment}
For \textsc{MetaTUS}, we evaluated combinations of topic embedding and topic filtering across all four modes described in the original work.
However, we report results only for the base mode, as the remaining variants (\textit{dtype}, \textit{dbpedia}, and \textit{dbpedia+dtype}) had higher runtimes without performance improvements.
These variants depend on mapping variables to external ontology classes, which proves ineffective in our setting.
The \textit{dtype} mode uses \textsc{Sherlock}~\cite{sherlock} to assign variables to one of 78 DBpedia classes via semantic type detection, while the \textit{dbpedia} mode employs an LLM for a similar mapping.
In both cases, the assigned DBpedia classes fail to capture the semantics of survey questionnaire variables.
As illustrated in \Cref{sherlock_example}, the classification of Q.37A as \emph{language} or Q.38A as \emph{region} do not reflect the underlying semantics of these variables.

\begin{figure}
    \begin{itemize}[leftmargin=0.1in,label=-]
        \footnotesize
        \item Q.37. WHAT WAS YOUR OCCUPATION FOUR YEARS AGO. I MEAN, WHAT KIND OF WORK WERE YOU DOING AROUND THE TIME OF THE 1948 ELECTION. \hfill \textbf{\textit{classification}}
        
        \item Q.37A. (IF R WAS EMPLOYED) WERE YOU WORKING FOR YOURSELF THEN, OR WERE YOU WORKING FOR SOMEONE ELSE. \hfill \textbf{\textit{language}}
        
        \item Q.38. (ASK ONLY IF R IS NOT HEAD OF HOUSEHOLD) WHAT KIND OF WORK WAS THE HEAD OF YOUR HOUSEHOLD DOING FOUR YEARS AGO, AROUND THE TIME OF THE 1948 ELECTION. \hfill \textbf{\textit{classification}}
        
        \item Q.38A. (IF HEAD WAS EMPLOYED) WAS HE WORKING FOR HIMSELF THEN, OR WAS HE WORKING FOR SOMEONE ELSE. \hfill \textbf{\textit{region}}
        
        \item Q.39. WHEN YOU WERE GOING TO GRADE SCHOOL OR HIGH SCHOOL, DID YOU GO TO A PAROCHIAL SCHOOL, A PRIVATE SCHOOL, OR A REGULAR PUBLIC SCHOOL. \hfill \textbf{\textit{type}}
        
        \item Q.40, 40A. WERE YOU OR THE HEAD OF YOUR HOUSEHOLD EVER IN THE ARMED SERVICES OF THE UNITED STATES. (IF YES) WHICH ONE OF YOU WAS THAT. \hfill \textbf{\textit{notes}}
        
        \item Q.40B. (IF YES TO REF.NO.234) WHAT YEARS DID THAT COVER. \hfill \textbf{\textit{notes}}
        
        \item FOR HEAD OF HOUSEHOLD (INCLUDING R IF HEAD) \hfill \textbf{\textit{language}}
        
        \item PARTY IDENTIFICATION AND STRENGTH \hfill \textbf{\textit{status}}
    \end{itemize}
    \caption{Examples of Semantic Type Prediction by Sherlock (used as a part of `dtype' mode) on LoStBench Variables:}
    \label{sherlock_example}
    \trimfigurespacing
\end{figure}

\paragraph{Analysis}

\begin{figure}[t]
    \centering
    
    % Top subfigure
    \begin{subfigure}[t]{\columnwidth}
        \centering
        \includegraphics[width=\linewidth]{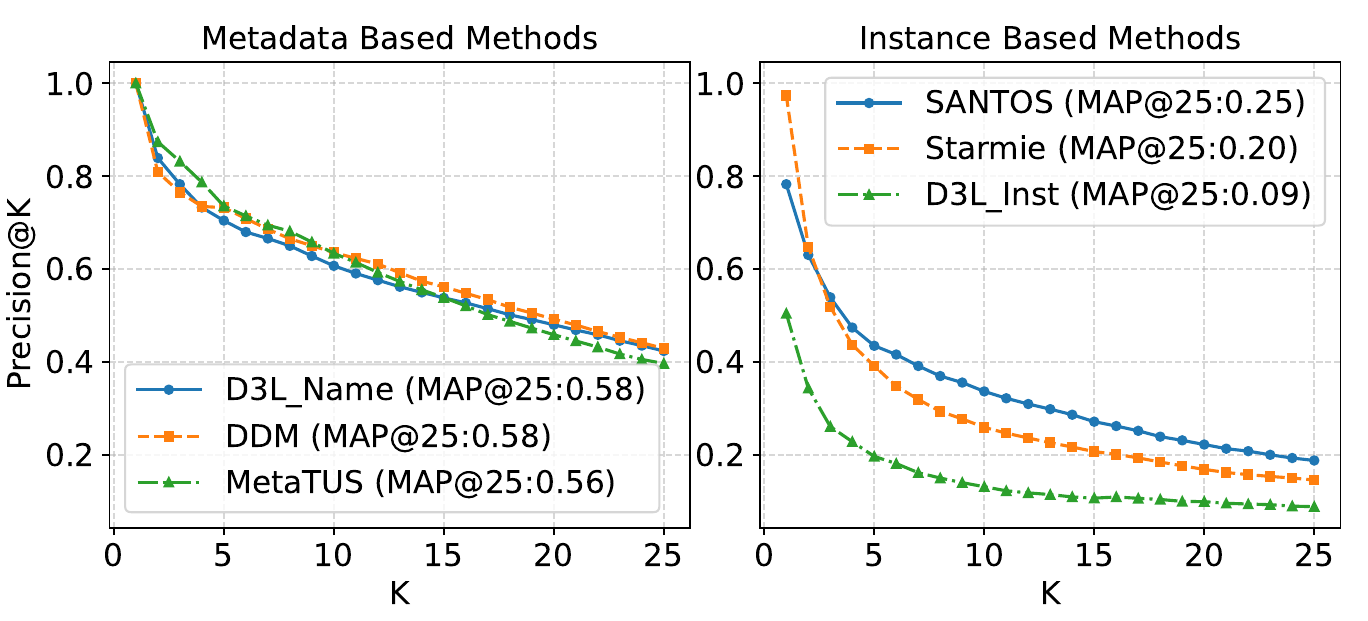}
        \caption{Precision@K Comparison}
        \label{fig:map@k}
    \end{subfigure}
    
    \vspace{0.1cm} 
    
    % Bottom subfigure
    \begin{subfigure}[t]{\columnwidth}
        \centering
        \includegraphics[width=\linewidth]{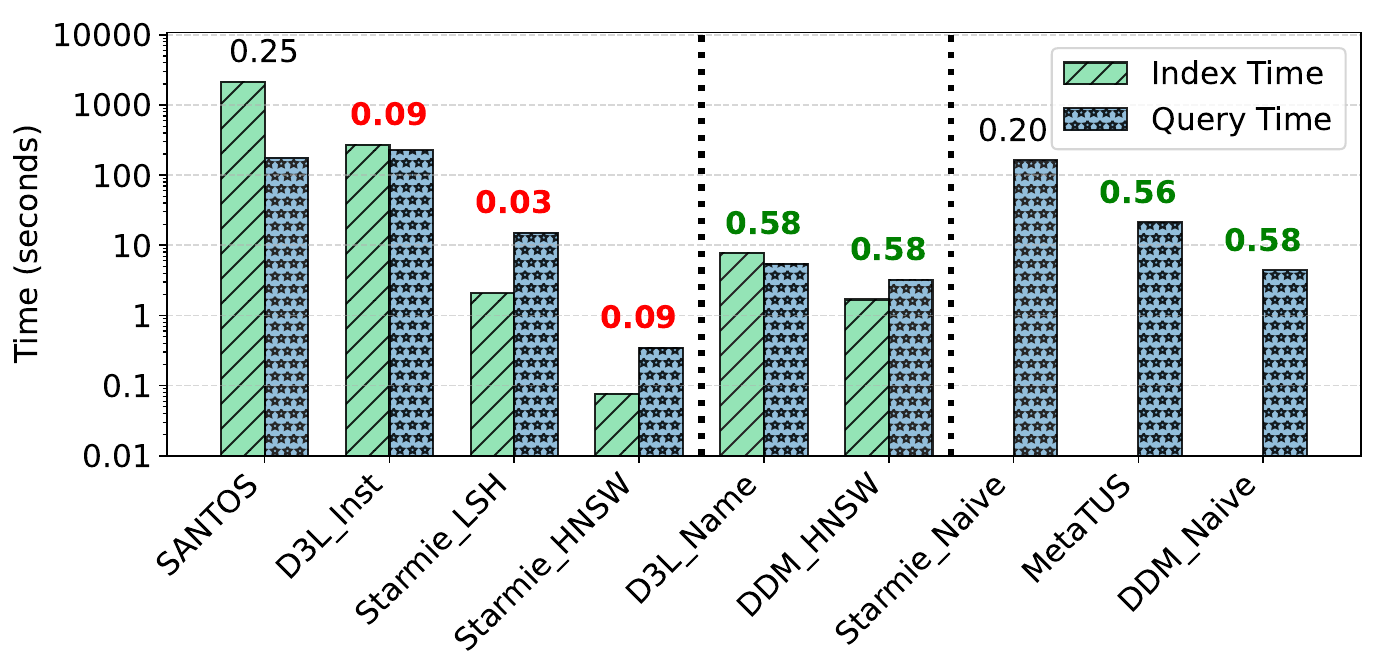}
        \caption{Runtimes for Indexing and Querying. Note.
        The three rightmost methods are naive search methods and do not include indexing time. Annotations at the top of each category represent their MAP@25.}
        \label{fig:tus_runtimes}
    \end{subfigure}
    
    \caption{Comparison of Effectiveness (Precision@K) and Efficiency (Query and Index times) for TUS methods.}
    \label{fig:tus_methods_comparison}
    \trimfigurespacing
\end{figure}

During indexing for \textsc{SANTOS}, entities for 1107 tables (approximately 80\%) could not be resolved from the YAGO \cite{yago_kb} knowledge base.
This is expected, as many literal values in the dataset do not correspond to entities in external knowledge bases.
Nevertheless, as shown in Figure~\ref{fig:map@k}, \textsc{SANTOS} achieves the highest Mean Average Precision@K among instance-based methods.
The MAP@K score reported in figure ~\ref{fig:map@k} for Starmie is for naive search.
The hnsw and lsh indexing search results obtained 50\% and 75\% worse results respectively as compared to the naive search counterpart.
Among metadata-based approaches, all three methods exhibit comparable performance in terms of precision@k.
However, in contrast to other benchmark datasets where SANTOS, Starmie, and MetaTUS often exceed 90\% MAP@K, performance on LoStBench is notably lower.
This answers \textbf{RQ \ref{q:category}} and further suggests that existing approaches struggle to capture the semantics of longitudinal survey data, highlighting the need for improved schema alignment models tailored to such datasets.

In terms of runtimes, the naive(brute-force) search for metadata-based methods (MetaTUS and DDM\_Naive) are substantially faster than the instance-based approach (Starmie\_Naive), as expected.
The LSH and HNSW based indexing variants of Starmie achieve the lowest query times among all methods;
however, this speedup comes at the cost of significantly reduced effectiveness.
In contrast, the HNSW-based variant of DDM not only provides a modest improvement in runtime over DDM\_Naive but also preserves MAP@K performance, despite relying on an approximate search mechanism.

% \end{enumerate}

%% file: src/conclusion.tex
\section{Conclusions}
\label{sec:conclusion}

We have presented LoStBench, a data integration benchmark for longitudinal studies based on the ANES longitudinal study dataset.
ANES differs from other datasets used in evaluating data integration in ways that include:
(i) Resistance to instance-based schema matching algorithms;
(ii) Strong attribute matching signal in the form of prose questions;
(iii) Hierarchical structure;
(iv) Very wide schemas; 
(v) Substantial redundancy; and 
(vi) Sequential structure in schema mappings.
We have demonstrated how existing algorithms for schema matching and related problems can be adapted to a longitudinal data integration task,
and evaluated a range of such algorithms applied in this new setting.
Notably, we find that metadata-based techniques outperform instance-based techniques on ANES.
We also find that algorithms like embedding-based matching, which can incorporate the prose text of questions, are able to outperform other schema matching techniques.
Finally, we find that techniques based on heavy-weight machine learning models are inconsistent in their ability to outperform lighter-weight models, and when they do, the resulting accuracy improvement does not justify the substantial increase in iteration time.

Through this work, we hope to motivate longitudinal studies as a compelling new use case for which new schema matching techniques may be developed.
LoStBench is available as a public github repository.

Although this work represents a first step into the space, many challenges remain open.
For example, surveys often involve redundancy to validate results; Not only does this result in many-to-many mappings, but also potential key violations.
A second, significant difference from classical schema matching is that some mappings (e.g., salary ranges) require transformation into a common frame of reference; a problem closer to data exchange.
As in data exchange, data values may be correlated, but not directly comparable.
For example, salary ranges from distinct years may overlap in large part, but not entirely.
Social longitudinal studies specifically exacerbate this problem, as even minor changes in question phrasing, interviewer, or survey context can bias the collected data for a given survey iteration.

%% file: main.bbl

\begin{thebibliography}{77}

%%% the Web address.

\ifx \showCODEN    \undefined \def \showCODEN     #1{\unskip}     \fi
\ifx \showDOI      \undefined \def \showDOI       #1{#1}\fi
\ifx \showISBNx    \undefined \def \showISBNx     #1{\unskip}     \fi
\ifx \showISBNxiii \undefined \def \showISBNxiii  #1{\unskip}     \fi
\ifx \showISSN     \undefined \def \showISSN      #1{\unskip}     \fi
\ifx \showLCCN     \undefined \def \showLCCN      #1{\unskip}     \fi
\ifx \shownote     \undefined \def \shownote      #1{#1}          \fi
\ifx \showarticletitle \undefined \def \showarticletitle #1{#1}   \fi
\ifx \showURL      \undefined \def \showURL       {\relax}        \fi
% invisible to TeX
\providecommand\bibfield[2]{#2}
\providecommand\bibinfo[2]{#2}
\providecommand\natexlab[1]{#1}
\providecommand\showeprint[2][]{arXiv:#2}

\bibitem[{American National Election Studies}(2026)]%
        {anes2026_codebook}
\bibfield{author}{\bibinfo{person}{{American National Election Studies}}.}
  \bibinfo{year}{2026}\natexlab{}.
\newblock \bibinfo{title}{\emph{ANES Time Series Cumulative Data File} [dataset
  and documentation]}.
\newblock
\newblock
\newblock
\shownote{February 5, 2026 version. \url{www.electionstudies.org}}.


\bibitem[Arenas et~al\mbox{.}(2014)]%
        {data_exchange_Marcelo_Arenas_textbook}
\bibfield{author}{\bibinfo{person}{Marcelo Arenas}, \bibinfo{person}{Pablo
  Barcel}, \bibinfo{person}{Leonid Libkin}, {and} \bibinfo{person}{Filip
  Murlak}.} \bibinfo{year}{2014}\natexlab{}.
\newblock \bibinfo{booktitle}{\emph{Foundations of Data Exchange}}.
\newblock \bibinfo{publisher}{Cambridge University Press},
  \bibinfo{address}{USA}.
\newblock
\showISBNx{1107016169}


\bibitem[Auer et~al\mbox{.}(2007)]%
        {dbpedia}
\bibfield{author}{\bibinfo{person}{S\"{o}ren Auer}, \bibinfo{person}{Christian
  Bizer}, \bibinfo{person}{Georgi Kobilarov}, \bibinfo{person}{Jens Lehmann},
  \bibinfo{person}{Richard Cyganiak}, {and} \bibinfo{person}{Zachary Ives}.}
  \bibinfo{year}{2007}\natexlab{}.
\newblock \showarticletitle{DBpedia: a nucleus for a web of open data}. In
  \bibinfo{booktitle}{\emph{Proceedings of the 6th International The Semantic
  Web and 2nd Asian Conference on Asian Semantic Web Conference}} (Busan,
  Korea) \emph{(\bibinfo{series}{ISWC'07/ASWC'07})}.
  \bibinfo{publisher}{Springer-Verlag}, \bibinfo{address}{Berlin, Heidelberg},
  \bibinfo{pages}{722–735}.
\newblock
\showISBNx{3540762973}


\bibitem[Aumueller et~al\mbox{.}(2005)]%
        {coma++}
\bibfield{author}{\bibinfo{person}{David Aumueller}, \bibinfo{person}{Hong-Hai
  Do}, \bibinfo{person}{Sabine Massmann}, {and} \bibinfo{person}{Erhard Rahm}.}
  \bibinfo{year}{2005}\natexlab{}.
\newblock \showarticletitle{Schema and ontology matching with COMA++}. In
  \bibinfo{booktitle}{\emph{Proceedings of the 2005 ACM SIGMOD International
  Conference on Management of Data}} (Baltimore, Maryland)
  \emph{(\bibinfo{series}{SIGMOD '05})}. \bibinfo{publisher}{Association for
  Computing Machinery}, \bibinfo{address}{New York, NY, USA},
  \bibinfo{pages}{906–908}.
\newblock
\showISBNx{1595930604}
\urldef\tempurl%
\url{https://doi.org/10.1145/1066157.1066283}
\showDOI{\tempurl}


\bibitem[Baxter et~al\mbox{.}(2003)]%
        {baxter2003blocking}
\bibfield{author}{\bibinfo{person}{Rohan Baxter}, \bibinfo{person}{Peter
  Christen}, {and} \bibinfo{person}{Tim Churches}.}
  \bibinfo{year}{2003}\natexlab{}.
\newblock \showarticletitle{A comparison of fast blocking methods for record
  linkage}. In \bibinfo{booktitle}{\emph{Workshop on Data Cleaning, Record
  Linkage and Object Consolidation}}. \bibinfo{pages}{25--27}.
\newblock


\bibitem[Bergamaschi et~al\mbox{.}(2001)]%
        {hetero_data_integration_sonia_bergamaschi}
\bibfield{author}{\bibinfo{person}{Sonia Bergamaschi}, \bibinfo{person}{Silvana
  Castano}, \bibinfo{person}{Maurizio Vincini}, {and} \bibinfo{person}{Domenico
  Beneventano}.} \bibinfo{year}{2001}\natexlab{}.
\newblock \showarticletitle{Semantic integration of heterogeneous information
  sources}.
\newblock \bibinfo{journal}{\emph{Data Knowl. Eng.}} \bibinfo{volume}{36},
  \bibinfo{number}{3} (\bibinfo{date}{March} \bibinfo{year}{2001}),
  \bibinfo{pages}{215–249}.
\newblock
\showISSN{0169-023X}
\urldef\tempurl%
\url{https://doi.org/10.1016/S0169-023X(00)00047-1}
\showDOI{\tempurl}


\bibitem[Bogatu et~al\mbox{.}(2020)]%
        {d3l_bogatu}
\bibfield{author}{\bibinfo{person}{Alex Bogatu}, \bibinfo{person}{Alvaro A.~A.
  Fernandes}, \bibinfo{person}{Norman~W. Paton}, {and}
  \bibinfo{person}{Nikolaos Konstantinou}.} \bibinfo{year}{2020}\natexlab{}.
\newblock \showarticletitle{Dataset Discovery in Data Lakes}. In
  \bibinfo{booktitle}{\emph{2020 IEEE 36th International Conference on Data
  Engineering (ICDE)}}. \bibinfo{publisher}{IEEE}, \bibinfo{pages}{709–720}.
\newblock
\urldef\tempurl%
\url{https://doi.org/10.1109/icde48307.2020.00067}
\showDOI{\tempurl}


\bibitem[Boubakir and Chaoui(2016)]%
        {boubakir2016pairwise_merge_multiple_models}
\bibfield{author}{\bibinfo{person}{Mohammed Boubakir} {and}
  \bibinfo{person}{Allaoua Chaoui}.} \bibinfo{year}{2016}\natexlab{}.
\newblock \showarticletitle{A Pairwise Approach for Model Merging}. In
  \bibinfo{booktitle}{\emph{Modelling and Implementation of Complex Systems}}.
  \bibinfo{publisher}{Springer}, \bibinfo{pages}{327--340}.
\newblock
\urldef\tempurl%
\url{https://doi.org/10.1007/978-3-319-30286-8_25}
\showDOI{\tempurl}


\bibitem[Boutaleb et~al\mbox{.}(2025)]%
        {boutaleb-etal-2025-somethings}
\bibfield{author}{\bibinfo{person}{Allaa Boutaleb}, \bibinfo{person}{Bernd
  Amann}, \bibinfo{person}{Hubert Naacke}, {and} \bibinfo{person}{Rafael
  Angarita}.} \bibinfo{year}{2025}\natexlab{}.
\newblock \showarticletitle{Something{'}s Fishy in the Data Lake: A Critical
  Re-evaluation of Table Union Search Benchmarks}. In
  \bibinfo{booktitle}{\emph{Proceedings of the 4th Table Representation
  Learning Workshop}}, \bibfield{editor}{\bibinfo{person}{Shuaichen Chang},
  \bibinfo{person}{Madelon Hulsebos}, \bibinfo{person}{Qian Liu},
  \bibinfo{person}{Wenhu Chen}, {and} \bibinfo{person}{Huan Sun}} (Eds.).
  \bibinfo{publisher}{Association for Computational Linguistics},
  \bibinfo{address}{Vienna, Austria}, \bibinfo{pages}{71--85}.
\newblock
\showISBNx{979-8-89176-268-8}
\urldef\tempurl%
\url{https://aclanthology.org/2025.trl-1.7/}
\showURL{%
\tempurl}


\bibitem[Buneman et~al\mbox{.}(1992)]%
        {buneman1992schema_merging_theoretical}
\bibfield{author}{\bibinfo{person}{Peter Buneman}, \bibinfo{person}{Susan~B.
  Davidson}, {and} \bibinfo{person}{Anthony Kosky}.}
  \bibinfo{year}{1992}\natexlab{}.
\newblock \showarticletitle{Theoretical Aspects of Schema Merging}. In
  \bibinfo{booktitle}{\emph{Proceedings of the 5th International Conference on
  Extending Database Technology (EDBT)}}. \bibinfo{publisher}{Springer},
  \bibinfo{pages}{152--167}.
\newblock
\urldef\tempurl%
\url{https://homepages.inf.ed.ac.uk/opb/papers/EDBT1992.pdf}
\showURL{%
\tempurl}


\bibitem[Carlson et~al\mbox{.}(2015)]%
        {poverty_panel_study}
\bibfield{author}{\bibinfo{person}{Deven Carlson}, \bibinfo{person}{Robert
  Haveman}, \bibinfo{person}{Barbara Wolfe}, \bibinfo{person}{Kyong~Wan Kang},
  {and} \bibinfo{person}{Hannah Miller}.} \bibinfo{year}{2015}\natexlab{}.
\newblock \bibinfo{booktitle}{\emph{Creating an Integrated Longitudinal Person
  and Household Dataset by Linking State-based Administrative Data}}.
\newblock \bibinfo{type}{{T}echnical {R}eport}. \bibinfo{institution}{Institute
  for Research on Poverty, University of Wisconsin--Madison}.
\newblock
\urldef\tempurl%
\url{https://www.neighborhoodindicators.org/sites/default/files/publications/SMR%20Progress%20report%2008%2025.pdf}
\showURL{%
\tempurl}
\newblock
\shownote{Accessed: 2026-04-01}.


\bibitem[Chelliah and de~Reuse(2011)]%
        {ChelliahDeReuse:2011}
\bibfield{author}{\bibinfo{person}{Shobhana~L. Chelliah} {and}
  \bibinfo{person}{Willem~J. de Reuse}.} \bibinfo{year}{2011}\natexlab{}.
\newblock \bibinfo{booktitle}{\emph{Handbook of descriptive linguistic
  fieldwork}}.
\newblock \bibinfo{publisher}{Dordrecht: Springer}.
\newblock


\bibitem[Chen et~al\mbox{.}(2026)]%
        {construm}
\bibfield{author}{\bibinfo{person}{Houming Chen}, \bibinfo{person}{Zhe Zhang},
  {and} \bibinfo{person}{H.~V. Jagadish}.} \bibinfo{year}{2026}\natexlab{}.
\newblock \bibinfo{title}{ConStruM: A Structure-Guided LLM Framework for
  Context-Aware Schema Matching}.
\newblock
\newblock
\showeprint[arxiv]{2601.20482}~[cs.DB]
\urldef\tempurl%
\url{https://arxiv.org/abs/2601.20482}
\showURL{%
\tempurl}


\bibitem[Chen et~al\mbox{.}(2009)]%
        {DBLP:conf/ictd/ChenCCDHP09}
\bibfield{author}{\bibinfo{person}{Kuang Chen}, \bibinfo{person}{Harr Chen},
  \bibinfo{person}{Neil Conway}, \bibinfo{person}{Heather Dolan},
  \bibinfo{person}{Joseph~M. Hellerstein}, {and} \bibinfo{person}{Tapan~S.
  Parikh}.} \bibinfo{year}{2009}\natexlab{}.
\newblock \showarticletitle{Improving data quality with dynamic forms}. In
  \bibinfo{booktitle}{\emph{{ICTD}}}. \bibinfo{publisher}{{IEEE}},
  \bibinfo{pages}{487}.
\newblock


\bibitem[Cong et~al\mbox{.}(2023)]%
        {pylon_tus}
\bibfield{author}{\bibinfo{person}{Tianji Cong}, \bibinfo{person}{Fatemeh
  Nargesian}, {and} \bibinfo{person}{HV Jagadish}.}
  \bibinfo{year}{2023}\natexlab{}.
\newblock \showarticletitle{Pylon: Semantic table union search in data lakes}.
\newblock \bibinfo{journal}{\emph{arXiv preprint arXiv:2301.04901}}
  (\bibinfo{year}{2023}).
\newblock


\bibitem[Daniel et~al\mbox{.}(2018)]%
        {Daniel2018}
\bibfield{author}{\bibinfo{person}{Hilary Daniel}, \bibinfo{person}{Sue~S.
  Bornstein}, {and} \bibinfo{person}{Gregory C.~Kane and}.}
  \bibinfo{year}{2018}\natexlab{}.
\newblock \showarticletitle{Addressing Social Determinants to Improve Patient
  Care and Promote Health Equity: {A}n {A}merican College of Physicians
  Position Paper}.
\newblock \bibinfo{journal}{\emph{Annals of Internal Medicine}}
  \bibinfo{volume}{168}, \bibinfo{number}{8} (\bibinfo{date}{April}
  \bibinfo{year}{2018}), \bibinfo{pages}{577}.
\newblock
\urldef\tempurl%
\url{https://doi.org/10.7326/m17-2441}
\showDOI{\tempurl}


\bibitem[Das~Sarma et~al\mbox{.}(2012)]%
        {finding_related_tables_alon_halevy}
\bibfield{author}{\bibinfo{person}{Anish Das~Sarma}, \bibinfo{person}{Lujun
  Fang}, \bibinfo{person}{Nitin Gupta}, \bibinfo{person}{Alon Halevy},
  \bibinfo{person}{Hongrae Lee}, \bibinfo{person}{Fei Wu},
  \bibinfo{person}{Reynold Xin}, {and} \bibinfo{person}{Cong Yu}.}
  \bibinfo{year}{2012}\natexlab{}.
\newblock \showarticletitle{Finding related tables}. In
  \bibinfo{booktitle}{\emph{Proceedings of the 2012 ACM SIGMOD International
  Conference on Management of Data}} (Scottsdale, Arizona, USA)
  \emph{(\bibinfo{series}{SIGMOD '12})}. \bibinfo{publisher}{Association for
  Computing Machinery}, \bibinfo{address}{New York, NY, USA},
  \bibinfo{pages}{817–828}.
\newblock
\showISBNx{9781450312479}
\urldef\tempurl%
\url{https://doi.org/10.1145/2213836.2213962}
\showDOI{\tempurl}


\bibitem[Deng et~al\mbox{.}(2020)]%
        {TURL}
\bibfield{author}{\bibinfo{person}{Xiang Deng}, \bibinfo{person}{Huan Sun},
  \bibinfo{person}{Alyssa Lees}, \bibinfo{person}{You Wu}, {and}
  \bibinfo{person}{Cong Yu}.} \bibinfo{year}{2020}\natexlab{}.
\newblock \bibinfo{title}{TURL: Table Understanding through Representation
  Learning}.
\newblock
\newblock
\showeprint[arxiv]{2006.14806}~[cs.IR]
\urldef\tempurl%
\url{https://arxiv.org/abs/2006.14806}
\showURL{%
\tempurl}


\bibitem[Deng et~al\mbox{.}(2024)]%
        {lakebench}
\bibfield{author}{\bibinfo{person}{Yuhao Deng}, \bibinfo{person}{Chengliang
  Chai}, \bibinfo{person}{Lei Cao}, \bibinfo{person}{Qin Yuan},
  \bibinfo{person}{Siyuan Chen}, \bibinfo{person}{Yanrui Yu},
  \bibinfo{person}{Zhaoze Sun}, \bibinfo{person}{Junyi Wang},
  \bibinfo{person}{Jiajun Li}, \bibinfo{person}{Ziqi Cao},
  \bibinfo{person}{Kaisen Jin}, \bibinfo{person}{Chi Zhang},
  \bibinfo{person}{Yuqing Jiang}, \bibinfo{person}{Yuanfang Zhang},
  \bibinfo{person}{Yuping Wang}, \bibinfo{person}{Ye Yuan},
  \bibinfo{person}{Guoren Wang}, {and} \bibinfo{person}{Nan Tang}.}
  \bibinfo{year}{2024}\natexlab{}.
\newblock \showarticletitle{LakeBench: A Benchmark for Discovering Joinable and
  Unionable Tables in Data Lakes}.
\newblock \bibinfo{journal}{\emph{Proc. VLDB Endow.}} \bibinfo{volume}{17},
  \bibinfo{number}{8} (\bibinfo{date}{April} \bibinfo{year}{2024}),
  \bibinfo{pages}{1925–1938}.
\newblock
\showISSN{2150-8097}
\urldef\tempurl%
\url{https://doi.org/10.14778/3659437.3659448}
\showDOI{\tempurl}


\bibitem[Do and Rahm(2002)]%
        {coma}
\bibfield{author}{\bibinfo{person}{Hong-Hai Do} {and} \bibinfo{person}{Erhard
  Rahm}.} \bibinfo{year}{2002}\natexlab{}.
\newblock \showarticletitle{COMA: a system for flexible combination of schema
  matching approaches}. In \bibinfo{booktitle}{\emph{Proceedings of the 28th
  International Conference on Very Large Data Bases}} (Hong Kong, China)
  \emph{(\bibinfo{series}{VLDB '02})}. \bibinfo{publisher}{VLDB Endowment},
  \bibinfo{pages}{610–621}.
\newblock


\bibitem[Doan et~al\mbox{.}(2001)]%
        {ml_for_sm}
\bibfield{author}{\bibinfo{person}{AnHai Doan}, \bibinfo{person}{Pedro
  Domingos}, {and} \bibinfo{person}{Alon~Y. Halevy}.}
  \bibinfo{year}{2001}\natexlab{}.
\newblock \showarticletitle{Reconciling schemas of disparate data sources: a
  machine-learning approach}.
\newblock \bibinfo{journal}{\emph{SIGMOD Rec.}} \bibinfo{volume}{30},
  \bibinfo{number}{2} (\bibinfo{date}{May} \bibinfo{year}{2001}),
  \bibinfo{pages}{509–520}.
\newblock
\showISSN{0163-5808}
\urldef\tempurl%
\url{https://doi.org/10.1145/376284.375731}
\showDOI{\tempurl}


\bibitem[Doan et~al\mbox{.}(2012)]%
        {data-integration_textbook}
\bibfield{author}{\bibinfo{person}{AnHai Doan}, \bibinfo{person}{Alon Halevy},
  {and} \bibinfo{person}{Zachary Ives}.} \bibinfo{year}{2012}\natexlab{}.
\newblock \bibinfo{booktitle}{\emph{Principles of Data Integration}
  (\bibinfo{edition}{1st} ed.)}.
\newblock \bibinfo{publisher}{Morgan Kaufmann Publishers Inc.},
  \bibinfo{address}{San Francisco, CA, USA}.
\newblock
\showISBNx{0124160441}


\bibitem[Dong et~al\mbox{.}(2021)]%
        {high_dim_join_discovery}
\bibfield{author}{\bibinfo{person}{Yuyang Dong}, \bibinfo{person}{Kunihiro
  Takeoka}, \bibinfo{person}{Chuan Xiao}, {and} \bibinfo{person}{Masafumi
  Oyamada}.} \bibinfo{year}{2021}\natexlab{}.
\newblock \bibinfo{title}{Efficient Joinable Table Discovery in Data Lakes: A
  High-Dimensional Similarity-Based Approach}.
\newblock
\newblock
\showeprint[arxiv]{2010.13273}~[cs.IR]
\urldef\tempurl%
\url{https://arxiv.org/abs/2010.13273}
\showURL{%
\tempurl}


\bibitem[Fan et~al\mbox{.}(2023)]%
        {5.fan}
\bibfield{author}{\bibinfo{person}{Grace Fan}, \bibinfo{person}{Jin Wang},
  \bibinfo{person}{Yuliang Li}, \bibinfo{person}{Dan Zhang}, {and}
  \bibinfo{person}{Renée Miller}.} \bibinfo{year}{2023}\natexlab{}.
\newblock \bibinfo{title}{Semantics-aware Dataset Discovery from Data Lakes
  with Contextualized Column-based Representation Learning}.
\newblock
\newblock
\showeprint[arxiv]{2210.01922}~[cs.DB]
\urldef\tempurl%
\url{https://arxiv.org/abs/2210.01922}
\showURL{%
\tempurl}


\bibitem[Fernandez et~al\mbox{.}(2018a)]%
        {aurum}
\bibfield{author}{\bibinfo{person}{Raul~Castro Fernandez},
  \bibinfo{person}{Ziawasch Abedjan}, \bibinfo{person}{Famien Koko},
  \bibinfo{person}{Gina Yuan}, \bibinfo{person}{Samuel Madden}, {and}
  \bibinfo{person}{Michael Stonebraker}.} \bibinfo{year}{2018}\natexlab{a}.
\newblock \showarticletitle{Aurum: {A} Data Discovery System}. In
  \bibinfo{booktitle}{\emph{ICDE}}. \bibinfo{pages}{1001--1012}.
\newblock


\bibitem[Fernandez et~al\mbox{.}(2018b)]%
        {semprop}
\bibfield{author}{\bibinfo{person}{Raul~Castro Fernandez},
  \bibinfo{person}{Essam Mansour}, \bibinfo{person}{Abdulhakim~Ali Qahtan},
  \bibinfo{person}{Ahmed~K. Elmagarmid}, \bibinfo{person}{Ihab~F. Ilyas},
  \bibinfo{person}{Samuel Madden}, \bibinfo{person}{Mourad Ouzzani},
  \bibinfo{person}{Michael Stonebraker}, {and} \bibinfo{person}{Nan Tang}.}
  \bibinfo{year}{2018}\natexlab{b}.
\newblock \showarticletitle{Seeping Semantics: Linking Datasets Using Word
  Embeddings for Data Discovery}.
\newblock \bibinfo{journal}{\emph{2018 IEEE 34th International Conference on
  Data Engineering (ICDE)}} (\bibinfo{year}{2018}), \bibinfo{pages}{989--1000}.
\newblock
\urldef\tempurl%
\url{https://api.semanticscholar.org/CorpusID:4210714}
\showURL{%
\tempurl}


\bibitem[Galea et~al\mbox{.}(2011)]%
        {Galea2011}
\bibfield{author}{\bibinfo{person}{Sandro Galea}, \bibinfo{person}{Melissa
  Tracy}, \bibinfo{person}{Katherine~J. Hoggatt}, \bibinfo{person}{Charles
  DiMaggio}, {and} \bibinfo{person}{Adam Karpati}.}
  \bibinfo{year}{2011}\natexlab{}.
\newblock \showarticletitle{Estimated Deaths Attributable to Social Factors in
  the {U}nited {S} tates}.
\newblock \bibinfo{journal}{\emph{American Journal of Public Health}}
  \bibinfo{volume}{101}, \bibinfo{number}{8} (\bibinfo{date}{Aug.}
  \bibinfo{year}{2011}), \bibinfo{pages}{1456--1465}.
\newblock
\urldef\tempurl%
\url{https://doi.org/10.2105/ajph.2010.300086}
\showDOI{\tempurl}


\bibitem[Good(2022)]%
        {Good:ScopeData}
\bibfield{author}{\bibinfo{person}{Jeff Good}.}
  \bibinfo{year}{2022}\natexlab{}.
\newblock \showarticletitle{The scope of linguistic data}.
\newblock In \bibinfo{booktitle}{\emph{The open handbook of linguistic data
  management}}, \bibfield{editor}{\bibinfo{person}{Andrea~L. Berez-Kroeker},
  \bibinfo{person}{Bradley McDonnell}, \bibinfo{person}{Eve Koller}, {and}
  \bibinfo{person}{Lauren~B. Collister}} (Eds.). \bibinfo{publisher}{Cambridge,
  MA: MIT Press}, \bibinfo{pages}{27--47}.
\newblock


\bibitem[Hulsebos et~al\mbox{.}(2019)]%
        {sherlock}
\bibfield{author}{\bibinfo{person}{Madelon Hulsebos}, \bibinfo{person}{Kevin
  Hu}, \bibinfo{person}{Michiel Bakker}, \bibinfo{person}{Emanuel Zgraggen},
  \bibinfo{person}{Arvind Satyanarayan}, \bibinfo{person}{Tim Kraska},
  \bibinfo{person}{{\c{ C}}a{\u{g}}atay Demiralp}, {and}
  \bibinfo{person}{C{\'e}sar Hidalgo}.} \bibinfo{year}{2019}\natexlab{}.
\newblock \showarticletitle{Sherlock: A Deep Learning Approach to Semantic Data
  Type Detection}. In \bibinfo{booktitle}{\emph{Proceedings of the 25th ACM
  SIGKDD International Conference on Knowledge Discovery \&\#38; Data Mining}}.
  \bibinfo{publisher}{ACM}.
\newblock


\bibitem[Hung et~al\mbox{.}(2014)]%
        {pay_as_you_go_sm}
\bibfield{author}{\bibinfo{person}{Nguyen Hung}, \bibinfo{person}{Tam Nguyen},
  \bibinfo{person}{Zoltan Miklos}, \bibinfo{person}{Karl Aberer},
  \bibinfo{person}{Avigdor Gal}, {and} \bibinfo{person}{Matthias Weidlich}.}
  \bibinfo{year}{2014}\natexlab{}.
\newblock \showarticletitle{Pay-as-you-go Reconciliation in Schema Matching
  Networks}.
\newblock \bibinfo{journal}{\emph{Proceedings - International Conference on
  Data Engineering}}, \bibinfo{pages}{220--231}.
\newblock
\showISBNx{978-1-4799-2555-1}
\urldef\tempurl%
\url{https://doi.org/10.1109/ICDE.2014.6816653}
\showDOI{\tempurl}


\bibitem[Iida et~al\mbox{.}(2021)]%
        {tabbie}
\bibfield{author}{\bibinfo{person}{Hiroshi Iida}, \bibinfo{person}{Dung Thai},
  \bibinfo{person}{Varun Manjunatha}, {and} \bibinfo{person}{Mohit Iyyer}.}
  \bibinfo{year}{2021}\natexlab{}.
\newblock \showarticletitle{{TABBIE}: Pretrained Representations of Tabular
  Data}. In \bibinfo{booktitle}{\emph{Proceedings of the 2021 Conference of the
  North American Chapter of the Association for Computational Linguistics:
  Human Language Technologies}}, \bibfield{editor}{\bibinfo{person}{Kristina
  Toutanova}, \bibinfo{person}{Anna Rumshisky}, \bibinfo{person}{Luke
  Zettlemoyer}, \bibinfo{person}{Dilek Hakkani-Tur},
  \bibinfo{person}{Iz~Beltagy}, \bibinfo{person}{Steven Bethard},
  \bibinfo{person}{Ryan Cotterell}, \bibinfo{person}{Tanmoy Chakraborty}, {and}
  \bibinfo{person}{Yichao Zhou}} (Eds.). \bibinfo{publisher}{Association for
  Computational Linguistics}, \bibinfo{address}{Online},
  \bibinfo{pages}{3446--3456}.
\newblock
\urldef\tempurl%
\url{https://doi.org/10.18653/v1/2021.naacl-main.270}
\showDOI{\tempurl}


\bibitem[Khatiwada et~al\mbox{.}(2023)]%
        {4.aamod}
\bibfield{author}{\bibinfo{person}{Aamod Khatiwada}, \bibinfo{person}{Grace
  Fan}, \bibinfo{person}{Roee Shraga}, \bibinfo{person}{Zixuan Chen},
  \bibinfo{person}{Wolfgang Gatterbauer}, \bibinfo{person}{Ren\'{e}e~J.
  Miller}, {and} \bibinfo{person}{Mirek Riedewald}.}
  \bibinfo{year}{2023}\natexlab{}.
\newblock \showarticletitle{SANTOS: Relationship-based Semantic Table Union
  Search}.
\newblock \bibinfo{journal}{\emph{Proc. ACM Manag. Data}} \bibinfo{volume}{1},
  \bibinfo{number}{1}, Article \bibinfo{articleno}{9} (\bibinfo{date}{May}
  \bibinfo{year}{2023}), \bibinfo{numpages}{25}~pages.
\newblock
\urldef\tempurl%
\url{https://doi.org/10.1145/3588689}
\showDOI{\tempurl}


\bibitem[Koutras et~al\mbox{.}(2020)]%
        {valentine}
\bibfield{author}{\bibinfo{person}{Christos Koutras}, \bibinfo{person}{George
  Siachamis}, \bibinfo{person}{Andra Ionescu}, \bibinfo{person}{Kyriakos
  Psarakis}, \bibinfo{person}{Jerry Brons}, \bibinfo{person}{Marios
  Fragkoulis}, \bibinfo{person}{Christoph Lofi}, \bibinfo{person}{Angela
  Bonifati}, {and} \bibinfo{person}{Asterios Katsifodimos}.}
  \bibinfo{year}{2020}\natexlab{}.
\newblock \showarticletitle{Valentine: Evaluating Matching Techniques for
  Dataset Discovery}.
\newblock \bibinfo{journal}{\emph{2021 IEEE 37th International Conference on
  Data Engineering (ICDE)}} (\bibinfo{year}{2020}), \bibinfo{pages}{468--479}.
\newblock
\urldef\tempurl%
\url{https://api.semanticscholar.org/CorpusID:222378204}
\showURL{%
\tempurl}


\bibitem[Langenecker et~al\mbox{.}(2023)]%
        {sportstables}
\bibfield{author}{\bibinfo{person}{Sven Langenecker},
  \bibinfo{person}{Christoph Sturm}, \bibinfo{person}{Christian Schalles},
  {and} \bibinfo{person}{Carsten Binnig}.} \bibinfo{year}{2023}\natexlab{}.
\newblock \showarticletitle{SportsTables: A New Corpus for Semantic Type
  Detection (Extended Version)}.
\newblock \bibinfo{journal}{\emph{Datenbank-Spektrum}}  \bibinfo{volume}{23}
  (\bibinfo{date}{10} \bibinfo{year}{2023}).
\newblock
\urldef\tempurl%
\url{https://doi.org/10.1007/s13222-023-00457-y}
\showDOI{\tempurl}


\bibitem[Langenecker et~al\mbox{.}(2024)]%
        {pythagoras}
\bibfield{author}{\bibinfo{person}{Sven Langenecker},
  \bibinfo{person}{Christoph Sturm}, \bibinfo{person}{Christian Schalles},
  {and} \bibinfo{person}{Carsten Binnig}.} \bibinfo{year}{2024}\natexlab{}.
\newblock \showarticletitle{Pythagoras: Semantic Type Detection of Numerical
  Data in Enterprise Data Lakes}. In \bibinfo{booktitle}{\emph{Proceedings of
  the 27th International Conference on Extending Database Technology (EDBT)}}
  (Paestum, Italy). \bibinfo{publisher}{OpenProceedings.org},
  \bibinfo{pages}{725--733}.
\newblock


\bibitem[Lenzerini(2002)]%
        {data_integration_theoretical_perspective}
\bibfield{author}{\bibinfo{person}{Maurizio Lenzerini}.}
  \bibinfo{year}{2002}\natexlab{}.
\newblock \showarticletitle{Data integration: a theoretical perspective}. In
  \bibinfo{booktitle}{\emph{Proceedings of the Twenty-First ACM
  SIGMOD-SIGACT-SIGART Symposium on Principles of Database Systems}} (Madison,
  Wisconsin) \emph{(\bibinfo{series}{PODS '02})}.
  \bibinfo{publisher}{Association for Computing Machinery},
  \bibinfo{address}{New York, NY, USA}, \bibinfo{pages}{233–246}.
\newblock
\showISBNx{1581135076}
\urldef\tempurl%
\url{https://doi.org/10.1145/543613.543644}
\showDOI{\tempurl}


\bibitem[Li and Tucker(2014)]%
        {long_cross_merge_get_pseudo_time_series}
\bibfield{author}{\bibinfo{person}{Yuanxi Li} {and} \bibinfo{person}{Allan
  Tucker}.} \bibinfo{year}{2014}\natexlab{}.
\newblock \showarticletitle{Integrating Clinical Data from Cross-Sectional and
  Longitudinal Studies}. In \bibinfo{booktitle}{\emph{2014 IEEE 27th
  International Symposium on Computer-Based Medical Systems}}.
  \bibinfo{pages}{465--466}.
\newblock
\urldef\tempurl%
\url{https://doi.org/10.1109/CBMS.2014.92}
\showDOI{\tempurl}


\bibitem[Liu et~al\mbox{.}(2025)]%
        {magneto}
\bibfield{author}{\bibinfo{person}{Yurong Liu}, \bibinfo{person}{Eduardo H.~M.
  Pena}, \bibinfo{person}{A\'{e}cio Santos}, \bibinfo{person}{Eden Wu}, {and}
  \bibinfo{person}{Juliana Freire}.} \bibinfo{year}{2025}\natexlab{}.
\newblock \showarticletitle{Magneto: Combining Small and Large Language Models
  for Schema Matching}.
\newblock \bibinfo{journal}{\emph{Proc. VLDB Endow.}} \bibinfo{volume}{18},
  \bibinfo{number}{8} (\bibinfo{date}{April} \bibinfo{year}{2025}),
  \bibinfo{pages}{2681–2694}.
\newblock
\showISSN{2150-8097}
\urldef\tempurl%
\url{https://doi.org/10.14778/3742728.3742757}
\showDOI{\tempurl}


\bibitem[Lugo-Martinez et~al\mbox{.}(2022)]%
        {long_study_growth_faltering_preterm_infants}
\bibfield{author}{\bibinfo{person}{Jose Lugo-Martinez}, \bibinfo{person}{Siwei
  Xu}, \bibinfo{person}{Justine Levesque}, \bibinfo{person}{Daniel Gallagher},
  \bibinfo{person}{Leslie~A. Parker}, \bibinfo{person}{Josef Neu},
  \bibinfo{person}{Christopher~J. Stewart}, \bibinfo{person}{Janet~E.
  Berrington}, \bibinfo{person}{Nicholas~D. Embleton}, \bibinfo{person}{Gregory
  Young}, \bibinfo{person}{Katherine~E. Gregory}, \bibinfo{person}{Misty Good},
  \bibinfo{person}{Arti Tandon}, \bibinfo{person}{David Genetti},
  \bibinfo{person}{Tracy Warren}, {and} \bibinfo{person}{Ziv Bar-Joseph}.}
  \bibinfo{year}{2022}\natexlab{}.
\newblock \showarticletitle{Integrating longitudinal clinical and microbiome
  data to predict growth faltering in preterm infants}.
\newblock \bibinfo{journal}{\emph{Journal of Biomedical Informatics}}
  \bibinfo{volume}{128} (\bibinfo{year}{2022}), \bibinfo{pages}{104031}.
\newblock
\showISSN{1532-0464}
\urldef\tempurl%
\url{https://doi.org/10.1016/j.jbi.2022.104031}
\showDOI{\tempurl}


\bibitem[Madhavan et~al\mbox{.}(2005)]%
        {corpus_based_sm_network}
\bibfield{author}{\bibinfo{person}{Jayant Madhavan}, \bibinfo{person}{Philip~A.
  Bernstein}, \bibinfo{person}{AnHai Doan}, {and} \bibinfo{person}{Alon
  Halevy}.} \bibinfo{year}{2005}\natexlab{}.
\newblock \showarticletitle{Corpus-Based Schema Matching}. In
  \bibinfo{booktitle}{\emph{Proceedings of the 21st International Conference on
  Data Engineering}} \emph{(\bibinfo{series}{ICDE '05})}.
  \bibinfo{publisher}{IEEE Computer Society}, \bibinfo{address}{USA},
  \bibinfo{pages}{57–68}.
\newblock
\showISBNx{0769522858}
\urldef\tempurl%
\url{https://doi.org/10.1109/ICDE.2005.39}
\showDOI{\tempurl}


\bibitem[Madhavan et~al\mbox{.}(2001)]%
        {cupid}
\bibfield{author}{\bibinfo{person}{Jayant Madhavan}, \bibinfo{person}{Philip~A.
  Bernstein}, {and} \bibinfo{person}{Erhard Rahm}.}
  \bibinfo{year}{2001}\natexlab{}.
\newblock \showarticletitle{Generic Schema Matching with Cupid}. In
  \bibinfo{booktitle}{\emph{Proceedings of the 27th International Conference on
  Very Large Data Bases}} \emph{(\bibinfo{series}{VLDB '01})}.
  \bibinfo{publisher}{Morgan Kaufmann Publishers Inc.}, \bibinfo{address}{San
  Francisco, CA, USA}, \bibinfo{pages}{49–58}.
\newblock
\showISBNx{1558608044}


\bibitem[Mandala(2017)]%
        {data_int_and_engg_general_article}
\bibfield{author}{\bibinfo{person}{Vishwanadham Mandala}.}
  \bibinfo{year}{2017}\natexlab{}.
\newblock \showarticletitle{Data Integration and Data Engineering Techniques}.
\newblock \bibinfo{journal}{\emph{International Journal of Scientific Research
  and Management (IJSRM)}} \bibinfo{volume}{5}, \bibinfo{number}{5}
  (\bibinfo{year}{2017}), \bibinfo{pages}{5354--5359}.
\newblock
\urldef\tempurl%
\url{https://doi.org/10.18535/ijsrm/v5i5.13}
\showDOI{\tempurl}


\bibitem[Manning et~al\mbox{.}(2008)]%
        {IR_textbook}
\bibfield{author}{\bibinfo{person}{Christopher~D. Manning},
  \bibinfo{person}{Prabhakar Raghavan}, {and} \bibinfo{person}{Hinrich
  Sch\"{u}tze}.} \bibinfo{year}{2008}\natexlab{}.
\newblock \bibinfo{booktitle}{\emph{Introduction to Information Retrieval}}.
\newblock \bibinfo{publisher}{Cambridge University Press},
  \bibinfo{address}{USA}.
\newblock
\showISBNx{0521865719}


\bibitem[Marcoulides and Grimm(2017)]%
        {measurement_scale_fusion}
\bibfield{author}{\bibinfo{person}{Katerina~M. Marcoulides} {and}
  \bibinfo{person}{Kevin~J. Grimm}.} \bibinfo{year}{2017}\natexlab{}.
\newblock \showarticletitle{Data Integration Approaches to Longitudinal Growth
  Modeling}.
\newblock \bibinfo{journal}{\emph{Educational and Psychological Measurement}}
  \bibinfo{volume}{77}, \bibinfo{number}{6} (\bibinfo{year}{2017}),
  \bibinfo{pages}{971--989}.
\newblock
\urldef\tempurl%
\url{https://doi.org/10.1177/0013164416664117}
\showDOI{\tempurl}
\showeprint{https://doi.org/10.1177/0013164416664117}


\bibitem[Markatou et~al\mbox{.}(2023)]%
        {10.3389/fmed.2023.1076794}
\bibfield{author}{\bibinfo{person}{Marianthi Markatou}, \bibinfo{person}{Oliver
  Kennedy}, \bibinfo{person}{Michael Brachmann}, \bibinfo{person}{Raktim
  Mukhopadhyay}, \bibinfo{person}{Arpan Dharia}, {and}
  \bibinfo{person}{Andrew~H. Talal}.} \bibinfo{year}{2023}\natexlab{}.
\newblock \showarticletitle{Social determinants of health derived from people
  with opioid use disorder: Improving data collection, integration and use with
  cross-domain collaboration and reproducible, data-centric, notebook-style
  workflows}.
\newblock \bibinfo{journal}{\emph{Frontiers in Medicine}}
  \bibinfo{volume}{Volume 10 - 2023} (\bibinfo{year}{2023}).
\newblock
\showISSN{2296-858X}
\urldef\tempurl%
\url{https://doi.org/10.3389/fmed.2023.1076794}
\showDOI{\tempurl}


\bibitem[Martorana et~al\mbox{.}(2022)]%
        {margherita_fair}
\bibfield{author}{\bibinfo{person}{Margherita Martorana},
  \bibinfo{person}{Tobias Kuhn}, \bibinfo{person}{Ronald Siebes}, {and}
  \bibinfo{person}{Jacco VanOssenbruggen}.} \bibinfo{year}{2022}\natexlab{}.
\newblock \showarticletitle{Aligning restricted access data with FAIR: a
  systematic review}.
\newblock \bibinfo{journal}{\emph{PeerJ Computer Science}} (\bibinfo{date}{Jul
  20} \bibinfo{year}{2022}).
\newblock
\urldef\tempurl%
\url{https://www.proquest.com/scholarly-journals/aligning-restricted-access-data-with-fair/docview/2691902429/se-2}
\showURL{%
\tempurl}
\newblock
\shownote{Copyright - © 2022 Martorana et al. This is an open access article
  distributed under the terms of the Creative Commons Attribution License:
  https://creativecommons.org/licenses/by/4.0/ (the “License”) , which
  permits unrestricted use, distribution, reproduction and adaptation in any
  medium and for any purpose provided that it is properly attributed. For
  attribution, the original author(s), title, publication source (PeerJ
  Computer Science) and either DOI or URL of the article must be cited.
  Notwithstanding the ProQuest Terms and Conditions, you may use this content
  in accordance with the terms of the License; Last updated - 2023-12-01}.


\bibitem[Martorana et~al\mbox{.}(2025)]%
        {2.martorana2025metadatadriventableunionsearch}
\bibfield{author}{\bibinfo{person}{Margherita Martorana},
  \bibinfo{person}{Tobias Kuhn}, {and} \bibinfo{person}{Jacco van
  Ossenbruggen}.} \bibinfo{year}{2025}\natexlab{}.
\newblock \bibinfo{title}{Metadata-driven Table Union Search: Leveraging
  Semantics for Restricted Access Data Integration}.
\newblock
\newblock
\showeprint[arxiv]{2502.20945}~[cs.DB]
\urldef\tempurl%
\url{https://arxiv.org/abs/2502.20945}
\showURL{%
\tempurl}


\bibitem[Melnik et~al\mbox{.}(2002)]%
        {sim_flooding}
\bibfield{author}{\bibinfo{person}{Sergey Melnik}, \bibinfo{person}{Hector
  Garcia-Molina}, {and} \bibinfo{person}{Erhard Rahm}.}
  \bibinfo{year}{2002}\natexlab{}.
\newblock \showarticletitle{Similarity Flooding: A Versatile Graph Matching
  Algorithm and Its Application to Schema Matching}. In
  \bibinfo{booktitle}{\emph{Proceedings of the 18th International Conference on
  Data Engineering}} \emph{(\bibinfo{series}{ICDE '02})}.
  \bibinfo{publisher}{IEEE Computer Society}, \bibinfo{address}{USA},
  \bibinfo{pages}{117}.
\newblock


\bibitem[Mitra et~al\mbox{.}(2018)]%
        {multiomics_obesity}
\bibfield{author}{\bibinfo{person}{Saptashwa Mitra}, \bibinfo{person}{Yu Qiu},
  \bibinfo{person}{Haley Moss}, \bibinfo{person}{Kaigang Li}, {and}
  \bibinfo{person}{Sangmi~Lee Pallickara}.} \bibinfo{year}{2018}\natexlab{}.
\newblock \showarticletitle{Effective Integration of Geotagged, Ancilliary
  Longitudinal Survey Datasets to Improve Adulthood Obesity Predictive Models}.
  In \bibinfo{booktitle}{\emph{2018 17th IEEE International Conference On
  Trust, Security And Privacy In Computing And Communications/ 12th IEEE
  International Conference On Big Data Science And Engineering
  (TrustCom/BigDataSE)}}. \bibinfo{pages}{1738--1746}.
\newblock
\urldef\tempurl%
\url{https://doi.org/10.1109/TrustCom/BigDataSE.2018.00263}
\showDOI{\tempurl}


\bibitem[Mugotitsa et~al\mbox{.}(2024)]%
        {mental_health_OMOP}
\bibfield{author}{\bibinfo{person}{Bylhah Mugotitsa},
  \bibinfo{person}{Tathagata Bhattacharjee}, \bibinfo{person}{Michael Ochola},
  \bibinfo{person}{Dorothy Mailosi}, \bibinfo{person}{David Amadi},
  \bibinfo{person}{Pauline Andeso}, \bibinfo{person}{Joseph Kuria},
  \bibinfo{person}{Reinpeter Momanyi}, \bibinfo{person}{Evans Omondi},
  \bibinfo{person}{Dan Kajungu}, \bibinfo{person}{Jim Todd},
  \bibinfo{person}{Agnes Kiragga}, {and} \bibinfo{person}{Jay Greenfield}.}
  \bibinfo{year}{2024}\natexlab{}.
\newblock \showarticletitle{Integrating longitudinal mental health data into a
  staging database: harnessing DDI-lifecycle and OMOP vocabularies within the
  INSPIRE Network Datahub}.
\newblock \bibinfo{journal}{\emph{Frontiers in Big Data}}
  \bibinfo{volume}{Volume 7 - 2024} (\bibinfo{year}{2024}).
\newblock
\showISSN{2624-909X}
\urldef\tempurl%
\url{https://doi.org/10.3389/fdata.2024.1435510}
\showDOI{\tempurl}


\bibitem[Nargesian et~al\mbox{.}(2018)]%
        {3.fatemeh}
\bibfield{author}{\bibinfo{person}{Fatemeh Nargesian}, \bibinfo{person}{Erkang
  Zhu}, \bibinfo{person}{Ken~Q. Pu}, {and} \bibinfo{person}{Ren \'{e}e~J.
  Miller}.} \bibinfo{year}{2018}\natexlab{}.
\newblock \showarticletitle{Table union search on open data}.
\newblock \bibinfo{journal}{\emph{Proc. VLDB Endow.}} \bibinfo{volume}{11},
  \bibinfo{number}{7} (\bibinfo{date}{March} \bibinfo{year}{2018}),
  \bibinfo{pages}{813–825}.
\newblock
\showISSN{2150-8097}
\urldef\tempurl%
\url{https://doi.org/10.14778/3192965.3192973}
\showDOI{\tempurl}


\bibitem[Nguyen(2024)]%
        {nguyen_paris_survival_analysis}
\bibfield{author}{\bibinfo{person}{van~Tuan Nguyen}.}
  \bibinfo{year}{2024}\natexlab{}.
\newblock \emph{\bibinfo{title}{{Integrating longitudinal data for enhanced
  survival analysis : methods and applications}}}.
\newblock Theses. \bibinfo{school}{{Universite Paris Cit{\'e}}}.
\newblock
\urldef\tempurl%
\url{https://theses.hal.science/tel-04903680}
\showURL{%
\tempurl}


\bibitem[Pal et~al\mbox{.}({[n.\,d.]})]%
        {alt_gen}
\bibfield{author}{\bibinfo{person}{Koyena Pal}, \bibinfo{person}{Aamod
  Khatiwada}, \bibinfo{person}{Roee Shraga}, {and} \bibinfo{person}{Ren{
  \'e}e~J Miller}.} \bibinfo{year}{[n.\,d.]}\natexlab{}.
\newblock \showarticletitle{ALT-GEN: Benchmarking Table Union Search using
  Large Language Models}.
\newblock \bibinfo{journal}{\emph{Proceedings of the VLDB Endowment. ISSN}}
  \bibinfo{volume}{2150} (\bibinfo{year}{[n.\,d.]}), \bibinfo{pages}{8097}.
\newblock


\bibitem[Pan et~al\mbox{.}(2021)]%
        {knowledge_graph_social_survey}
\bibfield{author}{\bibinfo{person}{Jeff~Z. Pan}, \bibinfo{person}{Elspeth
  Edelstein}, \bibinfo{person}{Patrik Bansky}, {and} \bibinfo{person}{Adam
  Wyner}.} \bibinfo{year}{2021}\natexlab{}.
\newblock \showarticletitle{A Knowledge Graph Based Approach to Social Science
  Surveys}.
\newblock \bibinfo{journal}{\emph{Data Intelligence}} \bibinfo{volume}{3},
  \bibinfo{number}{4} (\bibinfo{date}{10} \bibinfo{year}{2021}),
  \bibinfo{pages}{477--506}.
\newblock
\showISSN{2641-435X}
\urldef\tempurl%
\url{https://doi.org/10.1162/dint_a_00107}
\showDOI{\tempurl}


\bibitem[Parciak et~al\mbox{.}(2024)]%
        {sm_with_llm_monolithic}
\bibfield{author}{\bibinfo{person}{Marcel Parciak}, \bibinfo{person}{Brecht
  Vandevoort}, \bibinfo{person}{Frank Neven}, \bibinfo{person}{Liesbet~M.
  Peeters}, {and} \bibinfo{person}{Stijn Vansummeren}.}
  \bibinfo{year}{2024}\natexlab{}.
\newblock \bibinfo{title}{Schema Matching with Large Language Models: an
  Experimental Study}.
\newblock
\newblock
\showeprint[arxiv]{2407.11852}~[cs.DB]
\urldef\tempurl%
\url{https://arxiv.org/abs/2407.11852}
\showURL{%
\tempurl}


\bibitem[Pokharel et~al\mbox{.}(2024)]%
        {1.DDM}
\bibfield{author}{\bibinfo{person}{Pratik Pokharel}, \bibinfo{person}{Juseung
  Lee}, \bibinfo{person}{Oliver Kennedy}, \bibinfo{person}{Marianthi Markatou},
  \bibinfo{person}{Andrew Talal}, \bibinfo{person}{Jeff Good}, {and}
  \bibinfo{person}{Raktim Mukhopadhyay}.} \bibinfo{year}{2024}\natexlab{}.
\newblock \showarticletitle{Drag, Drop, Merge: A Tool for Streamlining
  Integration of Longitudinal Survey Instruments}. In
  \bibinfo{booktitle}{\emph{Proceedings of the 2024 Workshop on
  Human-In-the-Loop Data Analytics}} (Santiago, AA, Chile)
  \emph{(\bibinfo{series}{HILDA 24})}. \bibinfo{publisher}{Association for
  Computing Machinery}, \bibinfo{address}{New York, NY, USA},
  \bibinfo{pages}{1–7}.
\newblock
\showISBNx{9798400706936}
\urldef\tempurl%
\url{https://doi.org/10.1145/3665939.3665965}
\showDOI{\tempurl}


\bibitem[Pottinger and Bernstein(2003)]%
        {pottinger2003merging}
\bibfield{author}{\bibinfo{person}{Rachel Pottinger} {and}
  \bibinfo{person}{Philip~A. Bernstein}.} \bibinfo{year}{2003}\natexlab{}.
\newblock \showarticletitle{Merging Models Based on Given Correspondences}. In
  \bibinfo{booktitle}{\emph{Proceedings of the 29th International Conference on
  Very Large Data Bases (VLDB)}}. \bibinfo{publisher}{VLDB Endowment},
  \bibinfo{pages}{826--837}.
\newblock
\urldef\tempurl%
\url{https://www.vldb.org/conf/2003/papers/S26P01.pdf}
\showURL{%
\tempurl}


\bibitem[Quix et~al\mbox{.}(2007)]%
        {quix2007generic_merge}
\bibfield{author}{\bibinfo{person}{Christoph Quix}, \bibinfo{person}{David
  Kensche}, {and} \bibinfo{person}{Xiang Li}.} \bibinfo{year}{2007}\natexlab{}.
\newblock \showarticletitle{Generic Schema Merging}. In
  \bibinfo{booktitle}{\emph{Proceedings of the 19th International Conference on
  Advanced Information Systems Engineering (CAiSE)}}
  \emph{(\bibinfo{series}{Lecture Notes in Computer Science},
  Vol.~\bibinfo{volume}{4495})}. \bibinfo{publisher}{Springer},
  \bibinfo{pages}{127--141}.
\newblock
\urldef\tempurl%
\url{https://doi.org/10.1007/978-3-540-72988-4_10}
\showDOI{\tempurl}


\bibitem[Rahm and Bernstein(2001)]%
        {a_survey_of_sm_approaches}
\bibfield{author}{\bibinfo{person}{Erhard Rahm} {and}
  \bibinfo{person}{Philip~A. Bernstein}.} \bibinfo{year}{2001}\natexlab{}.
\newblock \showarticletitle{A survey of approaches to automatic schema
  matching}.
\newblock \bibinfo{journal}{\emph{The VLDB Journal}} \bibinfo{volume}{10},
  \bibinfo{number}{4} (\bibinfo{date}{Dec.} \bibinfo{year}{2001}),
  \bibinfo{pages}{334–350}.
\newblock
\showISSN{1066-8888}
\urldef\tempurl%
\url{https://doi.org/10.1007/s007780100057}
\showDOI{\tempurl}


\bibitem[Rebele et~al\mbox{.}(2016)]%
        {yago_kb}
\bibfield{author}{\bibinfo{person}{Thomas Rebele}, \bibinfo{person}{Fabian
  Suchanek}, \bibinfo{person}{Johannes Hoffart}, \bibinfo{person}{Joanna
  Biega}, \bibinfo{person}{Erdal Kuzey}, {and} \bibinfo{person}{Gerhard
  Weikum}.} \bibinfo{year}{2016}\natexlab{}.
\newblock \showarticletitle{YAGO: A Multilingual Knowledge Base from Wikipedia,
  Wordnet, and Geonames}. In \bibinfo{booktitle}{\emph{The Semantic Web –
  ISWC 2016: 15th International Semantic Web Conference, Kobe, Japan, October
  17–21, 2016, Proceedings, Part II}} (Kobe, Japan).
  \bibinfo{publisher}{Springer-Verlag}, \bibinfo{address}{Berlin, Heidelberg},
  \bibinfo{pages}{177–185}.
\newblock
\showISBNx{978-3-319-46546-3}
\urldef\tempurl%
\url{https://doi.org/10.1007/978-3-319-46547-0_19}
\showDOI{\tempurl}


\bibitem[Rodrigues and Silva(2021)]%
        {ml_techniques_for_sm_network}
\bibfield{author}{\bibinfo{person}{Diego Rodrigues} {and}
  \bibinfo{person}{Altigran Silva}.} \bibinfo{year}{2021}\natexlab{}.
\newblock \showarticletitle{A study on machine learning techniques for the
  schema matching network problem}.
\newblock \bibinfo{journal}{\emph{Journal of the Brazilian Computer Society}}
  \bibinfo{volume}{27} (\bibinfo{date}{12} \bibinfo{year}{2021}).
\newblock
\urldef\tempurl%
\url{https://doi.org/10.1186/s13173-021-00119-5}
\showDOI{\tempurl}


\bibitem[Scrocca et~al\mbox{.}(2021)]%
        {survey_ontology}
\bibfield{author}{\bibinfo{person}{Mario Scrocca}, \bibinfo{person}{Damiano
  Scandolari}, \bibinfo{person}{Gloria~Re Calegari}, \bibinfo{person}{Ilaria
  Baroni}, {and} \bibinfo{person}{Irene Celino}.}
  \bibinfo{year}{2021}\natexlab{}.
\newblock \showarticletitle{The Survey Ontology: Packaging Survey Research as
  Research Objects}. In \bibinfo{booktitle}{\emph{Proceedings of the 2nd
  Workshop on Data and Research Objects Management for Linked Open Science -
  co-located with ISWC 2021}}.
\newblock
\urldef\tempurl%
\url{https://doi.org/10.4126/FRL01-006429412,}
\showDOI{\tempurl}


\bibitem[Srinivas et~al\mbox{.}(2023)]%
        {lakebench_aamod}
\bibfield{author}{\bibinfo{person}{Kavitha Srinivas}, \bibinfo{person}{Julian
  Dolby}, \bibinfo{person}{Ibrahim Abdelaziz}, \bibinfo{person}{Oktie
  Hassanzadeh}, \bibinfo{person}{Harsha Kokel}, \bibinfo{person}{Aamod
  Khatiwada}, \bibinfo{person}{Tejaswini Pedapati}, \bibinfo{person}{Subhajit
  Chaudhury}, {and} \bibinfo{person}{Horst Samulowitz}.}
  \bibinfo{year}{2023}\natexlab{}.
\newblock \bibinfo{title}{LakeBench: Benchmarks for Data Discovery over Data
  Lakes}.
\newblock
\newblock
\showeprint[arxiv]{2307.04217}~[cs.DB]
\urldef\tempurl%
\url{https://arxiv.org/abs/2307.04217}
\showURL{%
\tempurl}


\bibitem[Su et~al\mbox{.}(2006)]%
        {hollistic_sm}
\bibfield{author}{\bibinfo{person}{Weifeng Su}, \bibinfo{person}{Jiying Wang},
  {and} \bibinfo{person}{Frederick Lochovsky}.}
  \bibinfo{year}{2006}\natexlab{}.
\newblock \showarticletitle{Holistic Schema Matching for Web Query Interfaces}.
  \bibinfo{pages}{77--94}.
\newblock
\showISBNx{978-3-540-32960-2}
\urldef\tempurl%
\url{https://doi.org/10.1007/11687238_8}
\showDOI{\tempurl}


\bibitem[Suhara et~al\mbox{.}(2022)]%
        {dosolo_doduo}
\bibfield{author}{\bibinfo{person}{Yoshihiko Suhara}, \bibinfo{person}{Jinfeng
  Li}, \bibinfo{person}{Yuliang Li}, \bibinfo{person}{Dan Zhang},
  \bibinfo{person}{\c{C}a\u{g}atay Demiralp}, \bibinfo{person}{Chen Chen},
  {and} \bibinfo{person}{Wang-Chiew Tan}.} \bibinfo{year}{2022}\natexlab{}.
\newblock \showarticletitle{Annotating Columns with Pre-trained Language
  Models}. In \bibinfo{booktitle}{\emph{Proceedings of the 2022 International
  Conference on Management of Data}} (Philadelphia, PA, USA)
  \emph{(\bibinfo{series}{SIGMOD '22})}. \bibinfo{publisher}{Association for
  Computing Machinery}, \bibinfo{address}{New York, NY, USA},
  \bibinfo{pages}{1493–1503}.
\newblock
\showISBNx{9781450392495}
\urldef\tempurl%
\url{https://doi.org/10.1145/3514221.3517906}
\showDOI{\tempurl}


\bibitem[Sun et~al\mbox{.}(2026)]%
        {tactus}
\bibfield{author}{\bibinfo{person}{Yongkang Sun}, \bibinfo{person}{Zhihao
  Ding}, \bibinfo{person}{Huiqiang Wang}, \bibinfo{person}{Reynold Cheng},
  {and} \bibinfo{person}{Jieming Shi}.} \bibinfo{year}{2026}\natexlab{}.
\newblock \bibinfo{title}{Efficient and Effective Table-Centric Table Union
  Search in Data Lakes}.
\newblock
\newblock
\showeprint[arxiv]{2603.17298}~[cs.DB]
\urldef\tempurl%
\url{https://arxiv.org/abs/2603.17298}
\showURL{%
\tempurl}


\bibitem[Susman et~al\mbox{.}(2025)]%
        {long_multimodels_separate_predictions_combinations}
\bibfield{author}{\bibinfo{person}{Aviad Susman}, \bibinfo{person}{Rupak
  Krishnamurthy}, \bibinfo{person}{Mohammad~Al Olaimat},
  \bibinfo{person}{Serdar Bozdag}, \bibinfo{person}{Bino Varghese},
  \bibinfo{person}{Nasim Sheikh-Bahaei}, {and} \bibinfo{person}{Gaurav
  Pandey}.} \bibinfo{year}{2025}\natexlab{}.
\newblock \showarticletitle{Longitudinal Ensemble Integration for Sequential
  Classification with Multimodal Data}. In \bibinfo{booktitle}{\emph{2025 IEEE
  International Conference on Digital Health (ICDH)}}. \bibinfo{pages}{31--40}.
\newblock
\urldef\tempurl%
\url{https://doi.org/10.1109/ICDH67620.2025.00014}
\showDOI{\tempurl}


\bibitem[Tasani and Yusof(2025)]%
        {di_challenges}
\bibfield{author}{\bibinfo{person}{Mimi~Haryani Tasani} {and}
  \bibinfo{person}{Mohd~Kamir Yusof}.} \bibinfo{year}{2025}\natexlab{}.
\newblock \showarticletitle{Challenges in Data Integration for Different Data
  Types: A Systematic Review}. In \bibinfo{booktitle}{\emph{2025 4th
  International Conference on Computer Technologies (ICCTech)}}.
  \bibinfo{pages}{98--102}.
\newblock
\urldef\tempurl%
\url{https://doi.org/10.1109/ICCTech66294.2025.00027}
\showDOI{\tempurl}


\bibitem[Taylor et~al\mbox{.}(2016)]%
        {Taylor2016}
\bibfield{author}{\bibinfo{person}{Lauren~A. Taylor},
  \bibinfo{person}{Annabel~Xulin Tan}, \bibinfo{person}{Caitlin~E. Coyle},
  \bibinfo{person}{Chima Ndumele}, \bibinfo{person}{Erika Rogan},
  \bibinfo{person}{Maureen Canavan}, \bibinfo{person}{Leslie~A. Curry}, {and}
  \bibinfo{person}{Elizabeth~H. Bradley}.} \bibinfo{year}{2016}\natexlab{}.
\newblock \showarticletitle{Leveraging the Social Determinants of Health: What
  Works?}
\newblock \bibinfo{journal}{\emph{{PLOS} {ONE}}} \bibinfo{volume}{11},
  \bibinfo{number}{8} (\bibinfo{date}{Aug.} \bibinfo{year}{2016}),
  \bibinfo{pages}{e0160217}.
\newblock
\urldef\tempurl%
\url{https://doi.org/10.1371/journal.pone.0160217}
\showDOI{\tempurl}


\bibitem[Thieberger and Berez(2012)]%
        {ThiebergerBerez:2012}
\bibfield{author}{\bibinfo{person}{Nicholas Thieberger} {and}
  \bibinfo{person}{Andrea~L. Berez}.} \bibinfo{year}{2012}\natexlab{}.
\newblock \showarticletitle{Linguistic data management}.
\newblock In \bibinfo{booktitle}{\emph{The {O}xford handbook of linguistic
  fieldwork}}, \bibfield{editor}{\bibinfo{person}{Nicholas Thieberger}} (Ed.).
  \bibinfo{publisher}{Oxford: OUP}, \bibinfo{pages}{90--118}.
\newblock


\bibitem[Thomas-Henkel and Schulman(2017)]%
        {ThomasHenkel2017}
\bibfield{author}{\bibinfo{person}{Caitlin Thomas-Henkel} {and}
  \bibinfo{person}{Meryl Schulman}.} \bibinfo{year}{2017}\natexlab{}.
\newblock \bibinfo{title}{Screening for Social Determinants of Health in
  Populations with Complex Needs: Implementation Considerations}.
\newblock
\newblock


\bibitem[Tu et~al\mbox{.}(2023)]%
        {unicorn}
\bibfield{author}{\bibinfo{person}{Jianhong Tu}, \bibinfo{person}{Ju Fan},
  \bibinfo{person}{Nan Tang}, \bibinfo{person}{Peng Wang},
  \bibinfo{person}{Guoliang Li}, \bibinfo{person}{Xiaoyong Du},
  \bibinfo{person}{Xiaofeng Jia}, {and} \bibinfo{person}{Song Gao}.}
  \bibinfo{year}{2023}\natexlab{}.
\newblock \showarticletitle{Unicorn: A Unified Multi-tasking Model for
  Supporting Matching Tasks in Data Integration}.
\newblock \bibinfo{journal}{\emph{Proc. ACM Manag. Data}} \bibinfo{volume}{1},
  \bibinfo{number}{1}, Article \bibinfo{articleno}{84} (\bibinfo{date}{May}
  \bibinfo{year}{2023}), \bibinfo{numpages}{26}~pages.
\newblock
\urldef\tempurl%
\url{https://doi.org/10.1145/3588938}
\showDOI{\tempurl}


\bibitem[Yakout et~al\mbox{.}(2012)]%
        {infogather}
\bibfield{author}{\bibinfo{person}{Mohamed Yakout}, \bibinfo{person}{Kris
  Ganjam}, \bibinfo{person}{Kaushik Chakrabarti}, {and}
  \bibinfo{person}{Surajit Chaudhuri}.} \bibinfo{year}{2012}\natexlab{}.
\newblock \showarticletitle{InfoGather: entity augmentation and attribute
  discovery by holistic matching with web tables}. In
  \bibinfo{booktitle}{\emph{Proceedings of the 2012 ACM SIGMOD International
  Conference on Management of Data}} (Scottsdale, Arizona, USA)
  \emph{(\bibinfo{series}{SIGMOD '12})}. \bibinfo{publisher}{Association for
  Computing Machinery}, \bibinfo{address}{New York, NY, USA},
  \bibinfo{pages}{97–108}.
\newblock
\showISBNx{9781450312479}
\urldef\tempurl%
\url{https://doi.org/10.1145/2213836.2213848}
\showDOI{\tempurl}


\bibitem[Zhang et~al\mbox{.}(2020)]%
        {sato}
\bibfield{author}{\bibinfo{person}{Dan Zhang}, \bibinfo{person}{Yoshihiko
  Suhara}, \bibinfo{person}{Jinfeng Li}, \bibinfo{person}{Madelon Hulsebos},
  \bibinfo{person}{{\c{C}}a{\u{g}}atay Demiralp}, {and}
  \bibinfo{person}{Wang-Chiew Tan}.} \bibinfo{year}{2020}\natexlab{}.
\newblock \showarticletitle{Sato: Contextual Semantic Type Detection in
  Tables}.
\newblock \bibinfo{journal}{\emph{Proc. VLDB Endow.}} \bibinfo{volume}{13},
  \bibinfo{number}{12} (\bibinfo{year}{2020}), \bibinfo{pages}{1835–1848}.
\newblock
\urldef\tempurl%
\url{https://doi.org/10.14778/3407790.3407793}
\showURL{%
\tempurl}


\bibitem[Zhang et~al\mbox{.}(2011)]%
        {distr_based}
\bibfield{author}{\bibinfo{person}{Meihui Zhang}, \bibinfo{person}{Marios
  Hadjieleftheriou}, \bibinfo{person}{Beng~Chin Ooi},
  \bibinfo{person}{Cecilia~M. Procopiuc}, {and} \bibinfo{person}{Divesh
  Srivastava}.} \bibinfo{year}{2011}\natexlab{}.
\newblock \showarticletitle{Automatic discovery of attributes in relational
  databases}. In \bibinfo{booktitle}{\emph{Proceedings of the 2011 ACM SIGMOD
  International Conference on Management of Data}} (Athens, Greece)
  \emph{(\bibinfo{series}{SIGMOD '11})}. \bibinfo{publisher}{Association for
  Computing Machinery}, \bibinfo{address}{New York, NY, USA},
  \bibinfo{pages}{109–120}.
\newblock
\showISBNx{9781450306614}
\urldef\tempurl%
\url{https://doi.org/10.1145/1989323.1989336}
\showDOI{\tempurl}


\bibitem[Zhang et~al\mbox{.}(2024)]%
        {nested_data_search}
\bibfield{author}{\bibinfo{person}{Yi Zhang}, \bibinfo{person}{Peter~Baile
  Chen}, {and} \bibinfo{person}{Zachary~G. Ives}.}
  \bibinfo{year}{2024}\natexlab{}.
\newblock \showarticletitle{Searching Data Lakes for Nested and Joined Data}.
\newblock \bibinfo{journal}{\emph{Proc. VLDB Endow.}} \bibinfo{volume}{17},
  \bibinfo{number}{11} (\bibinfo{date}{July} \bibinfo{year}{2024}),
  \bibinfo{pages}{3346–3359}.
\newblock
\showISSN{2150-8097}
\urldef\tempurl%
\url{https://doi.org/10.14778/3681954.3682005}
\showDOI{\tempurl}


\bibitem[Zhu et~al\mbox{.}(2019)]%
        {josie}
\bibfield{author}{\bibinfo{person}{Erkang Zhu}, \bibinfo{person}{Dong Deng},
  \bibinfo{person}{Fatemeh Nargesian}, {and} \bibinfo{person}{Ren \'{e}e~J.
  Miller}.} \bibinfo{year}{2019}\natexlab{}.
\newblock \showarticletitle{JOSIE: Overlap Set Similarity Search for Finding
  Joinable Tables in Data Lakes}. In \bibinfo{booktitle}{\emph{Proceedings of
  the 2019 International Conference on Management of Data}} (Amsterdam,
  Netherlands) \emph{(\bibinfo{series}{SIGMOD '19})}.
  \bibinfo{publisher}{Association for Computing Machinery},
  \bibinfo{address}{New York, NY, USA}, \bibinfo{pages}{847–864}.
\newblock
\showISBNx{9781450356435}
\urldef\tempurl%
\url{https://doi.org/10.1145/3299869.3300065}
\showDOI{\tempurl}


\end{thebibliography}
